\documentclass[twocolumn]{aastex631}

\defcitealias{Michail2021c}{Paper I}
\newcommand{\Sp}{{\it Spitzer\/}}
\newcommand{\Ch}{{\it Chandra\/}}

\usepackage{graphicx}	
\usepackage{amsmath}	
\usepackage{amssymb}	

\received{}
\revised{}
\accepted{}

\submitjournal{ApJ}

\begin{document}

\title{Multiwavelength Observations of Sgr A*. II. 2019 July 21 and 26}
\shorttitle{Michail et al.}


\author[0000-0003-3503-3446]{Joseph M. Michail}
\affiliation{Department of Physics and Astronomy, Northwestern University, 2145 Sheridan Rd, Evanston, IL 60208, USA}
\affiliation{Center for Interdisciplinary Exploration and Research in Astrophysics (CIERA), Northwestern University, 1800 Sherman Ave, Evanston, IL 60201, USA}

\author{Farhad Yusef-Zadeh}
\affiliation{Department of Physics and Astronomy, Northwestern University, 2145 Sheridan Rd, Evanston, IL 60208, USA}
\affiliation{Center for Interdisciplinary Exploration and Research in Astrophysics (CIERA), Northwestern University, 1800 Sherman Ave, Evanston, IL 60201, USA}

\author[0000-0002-1737-0871]{Mark Wardle}
\affiliation{Astrophysics and Space Technologies Research Centre and School of Mathematics and Physical Sciences, Macquarie University, Sydney, NSW 2109, Australia}

\author[0000-0002-1568-579X]{Devaky Kunneriath}
\affiliation{National Radio Astronomy Observatory, 520 Edgemont Road, Charlottesville, VA 22903, USA}

\author[0000-0002-5599-4650]{Joseph L. Hora}
\affiliation{Center for Astrophysics $|$ Harvard \& Smithsonian, 60 Garden Street, MS-65, Cambridge, MA 02138-1516, USA}

\author[0000-0001-6664-7585]{Howard Bushouse}
\affiliation{Space Telescope Science Institute, Baltimore, MD 21218, USA}

\author[0000-0002-0670-0708]{Giovanni G. Fazio}
\affiliation{Center for Astrophysics $|$ Harvard \& Smithsonian, 60 Garden Street, MS-65, Cambridge, MA 02138-1516, USA}

\author[0000-0001-9564-0876]{Sera Markoff}
\affiliation{Anton Pannekoek Institute for Astronomy, University of Amsterdam, Science Park 904, 1098 XH, Amsterdam, The Netherlands}
\affiliation{Gravitation and Astroparticle Physics Amsterdam (GRAPPA) Institute, University of Amsterdam, Science Park 904, 1098 XH Amsterdam, The Netherlands}

\author{Howard A. Smith}
\affiliation{Center for Astrophysics $|$ Harvard \& Smithsonian, 60 Garden Street, MS-65, Cambridge, MA 02138-1516, USA}

\correspondingauthor{Joseph M. Michail}
\email{michail@u.northwestern.edu}

\begin{abstract}
    We report on the final two days of a multiwavelength campaign of Sgr A* observing in the radio, submillimeter, infrared, and X-ray bands in July 2019. Sgr A* was remarkably active, showing multiple flaring events across the electromagnetic spectrum. We detect a transient $\sim35$-minute periodicity feature in \Sp\ light curves on 21 July 2019. Time-delayed emission was detected in ALMA light curves, suggesting a hotspot within the accretion flow on a stable orbit. On the same night, we observe a decreased flux in the submillimeter light curve following an X-ray flare detected by \Ch\ and model the feature with an adiabatically expanding synchrotron hotspot occulting the accretion flow. The event is produced by a plasma $0.55~R_{\text{S}}$ in radius with an electron spectrum $p=2.84$. It is threaded by a $\sim130$ Gauss magnetic field and expands at $0.6\%$ the speed of light. Finally, we reveal an unambiguous flare in the infrared, submillimeter, and radio, demonstrating that the variable emission is intrinsically linked. We jointly fit the radio and submillimeter light curves using an adiabatically expanding synchrotron hotspot and find it is produced by a plasma with an electron spectrum $p=0.59$, $187$ Gauss magnetic field, and radius $0.47~R_{\text{S}}$ that expands at $0.029c$. In both cases, the uncertainty in the appropriate lower and upper electron energy bounds may inflate the derived equipartition field strengths by a factor of 2 or more. Our results confirm that both synchrotron- and adiabatic-cooling processes are involved in the variable emission's evolution at submillimeter and infrared wavelengths.
\end{abstract}



\section{Introduction}
    The Milky Way Galaxy hosts Sagittarius A* (Sgr A*), a $4.152\times10^{6}~\text{M}_\odot$ supermassive black hole, at its center. At a distance of $8178$ pc \citep{Boehle2016, Gravity2019}, it is one of only two black holes whose angular extent is large enough to be directly imaged by Earth-sized very long baseline interferometry (VLBI) experiments like the Event Horizon Telescope \citep[EHT; ][]{EHT2022a}. One complication in imaging its near-event-horizon-scale features is its variability. Sgr A* has a gravitational timescale $t_g = r_{g}/c = GM/c^3\sim20$ seconds, and the Keplerian orbital period at the innermost stable circular orbit (ISCO) is $\sim30$ minutes if Sgr A* does not spin \citep{Gravity2018}, making it variable over a short duration. Imaging near-event-horizon features requires observations many times longer than these variability timescales, which corrupts the visibilities produced by intrinsic structures and smears them in the final image.  Statistical attempts have been used to mitigate some of its variable nature and create a best-fit event-horizon-scale image \citep{Broderick2022}. However, understanding the cause of the variability will ultimately benefit future imaging attempts of this source.

    Sgr A*'s variability has been observed in the radio \citep[i.e.,][]{Lu2011, FYZ2011, Brinkerink2015}, submillimeter/millimeter \citep[submm/mm; i.e.,][]{Dexter2014, Subroweit2017, Iwata2020}, infrared \citep[IR; i.e.,][]{Hora2014, Witzel2018}, and X-rays \citep[i.e.,][]{Baganoff2001, Markoff2001, Neilsen2015,Ponti2015,Yuan2016}. Large-amplitude hourly timescale variability, which we denote ``flares,'' occurs roughly once to a few times per day. At radio and (sub)mm, these flares appear in addition to slowly varying ``quiescent''-like emission from an accretion flow near the event horizon. At radio frequencies, this emission is optically thick. Toward submm frequencies, its optical depth decreases, and the emission is either marginally optically thick or optically thin at 230 GHz and optically thin at 345 GHz \citep[i.e.,][]{Bower2019}. \citet{Wielgus2022a} suggest that the submm quiescent emission at 230 GHz and higher is optically thin, but there may be regions within it that have higher-than-average optical depth. The causes of the flares and the slowly varying quiescent emission are unknown; multiwavelength observations are crucial to unraveling this mystery.
    

    Coordinated multiwavelength observations have led to insights into the flaring emission. The time of the peak flaring emission at lower frequencies is delayed relative to those at high frequencies. Between radio and submm, the time delays are on the order of tens of minutes to an hour \citep[i.e.,][]{Brinkerink2015, Mossoux2017, Michail2021c}. Similar lags have been discovered across a single radio frequency band using the NSF's Karl G. Jansky Very Large Array \citep[VLA;][]{Brinkerink2021, Michail2021b}. At IR, the emission has been observed to lead or simultaneously occur with submm \citep[i.e.,][]{YusefZadeh2006a, Fazio2018}. The appearance or lack of delays depends on the observing frequency (i.e., 230 vs. 345 GHz) and the emission's physical properties. 
    
    There is a curious connection between flares in the IR and X-rays. First, every X-ray flare has an IR counterpart, but the opposite is not true \citep{DoddsEden2009}. These ``missing'' flares may be caused by limited observational sensitivity in the X-rays, as there is large-scale quiescent X-ray emission near Sgr A*.  The missing X-ray flares may not be sufficiently bright to overcome this large-scale emission, leaving only the strongest flares to be observed. Second, the time delay between IR and X-ray light curves has not been well-established. \citet{Boyce2019} suggested the X-ray might lead the IR by 10-20 minutes after analyzing more than 100 hours of coordinated IR and X-ray monitoring. This ensemble-average value is only significant at a 68\% confidence interval. The authors conclude additional data are required to constrain the correlation. 

    One model that attempts to explain the connection between IR, submm, and radio flaring is the adiabatically-expanding synchrotron hotspot model \citep{VDL1966, FYZ2006}. It describes the time- and frequency-dependent evolution of an initially optically-thick synchrotron-emitting plasma as it expands. The time delay naturally explains the dynamic expansion as the emission transitions from optically thick to thin. This picture predicts unique continuum and polarimetric light curves, which depend on properties such as the embedded magnetic field, electron spectrum, and expansion speed. In a series of earlier papers, we explored the multiwavelength and polarimetric nature of these hotspots. \citet{Michail2023} tested this picture using a full-radiative transfer prescription to model 235 GHz full-polarization light curves of a detected flare from ALMA. For the first time, we constrained the three-dimensional orientation of the embedded magnetic field. In \citet[][further referred to as \citetalias{Michail2021c}]{Michail2021c}, we used this picture to connect the IR and submm variability and predict the resulting radio emission. 
    
    In July 2019, we participated in organizing a ground- and space-based campaign to simultaneously monitor Sgr A* from radio to X-ray energies. Several papers have analyzed these data but do not focus on the complementary radio and Hubble Space Telescope (HST) observations. \citet{Abuter2021} analyzed a remarkable flare on 18 July 2019, which occurred only a few months after the brightest-ever IR flare with a dereddened $2.15~\mu$m peak flux of $\sim60$ mJy \citep[][]{Do2019}. Using IR (\Sp\ \textit{Space Telescope} and GRAVITY) and X-ray (\Ch\ \textit{X-ray Observatory} and NuSTAR) observations, they show its mean and time-resolved spectral energy distributions are consistent with synchrotron self-Compton or a pure synchrotron spectrum with a cooling break. The synchrotron self-Compton model was unable to reproduce any significant submm emission (first reported in \citetalias{Michail2021c}), while the latter produced submm emission at levels consistent with the historic average. However, properties of this model, such as higher-than-inferred electron densities in the accretion flow, were drawbacks. \citet{Boyce2022} reported the remaining two days of ALMA, \Sp, and \Ch\ light curves from the July 2019 campaign and remodeled the first day's results to include the ALMA data using an adiabatically-expanding source region. Their models rejected the optically-thin case reported in \citetalias{Michail2021c} where both the submm and IR flares were produced by a single synchrotron component, as the IR spectral index was not consistent with the observations and could not reproduce the X-ray emission using synchrotron self-Compton. The emission was consistent with an adiabatically-expanding model such that the IR and X-ray emission were both produced by synchrotron self-Compton processes and the submm was produced by synchrotron only. However, such a model predicted similarly high electron densities ($n_e\sim10^{9-10}$ cm$^{-3}$) like those modeled by \citet{Abuter2021}. 
    
    \citet{Boyce2022} focused only on modeling multiwavelength features on the first day of observation and did not include contemporaneous HST (IR) and VLA (radio) observations during this campaign. Here, we present the remaining two days of monitoring with all participating observatories. We discuss data reduction and calibration for each instrument in Section \ref{sec:data}. In Section \ref{sec:analysis}, we present the time delay analysis between each pair of instruments and the periodogram analysis of the IR data. In Section \ref{sec:discussion}, we use the adiabatic-expansion picture (described in Section \ref{sec:models}) to model the detected features and summarize our conclusions in Section \ref{sec:conclusions}. Our adopted mass of Sgr A* throughout this paper corresponds to a Schwarzschild radius $R_{\text{S}} = 1.23\times10^{12}$ cm.
    
\section{Observations and Data Reduction}\label{sec:data}
    Here, we describe the observing details and calibration procedures for the data used in this analysis. \citetalias{Michail2021c} explained the techniques in greater detail for the ALMA, \Sp, and \Ch\ data; we give summaries for these instruments below. We present detailed explanations for the VLA and HST data.
    
    \subsection{Radio: Very Large Array}\label{ssec:vla}
        The VLA observed Sgr A* on three days: 18 July 2019 (Project ID: 19A-490, PI: Yusef-Zadeh), 21 July 2019, and 26 July 2019 (Project ID: 19A-229, PI: Yusef-Zadeh). Observations on 18 July 2019 were completed in the hybrid BnA array at Q-band (43 GHz). On the 21st and 26th, the VLA was in the A configuration and used the Q and Ka (35 GHz) bands, respectively. The 3-bit correlator was used to observe over 8 GHz bandwidth and is composed of 63 intermediate frequencies (IFs), each with a bandwidth of 128 MHz. Every IF is subdivided into 64 2-MHz wide channels recording full-polarization products.
        
        Observations on 18 and 21 July were not successfully completed owing to several compounding issues. The cycle times between Sgr A* and the phase calibrator were lengthier than required in the most extended configurations ($\sim20$ minutes used versus the recommended $\sim2$ minutes). Additionally, the pointing calibrations were not completed often enough for high-frequency observing, nominally at a cadence of $45-60$ minutes. Finally, strong winds above the specifications for Q-band observing occurred during these two nights. Despite our best efforts, these data were not salvageable to extract light curves. While the 26 July observations were set up similarly, better weather conditions and the wider primary beam at Ka-band reduced many of the pointing issues. Therefore, we only make use of the Ka-band observations in this analysis.
        
\begin{figure}
            \centering
            \includegraphics[width=\columnwidth]{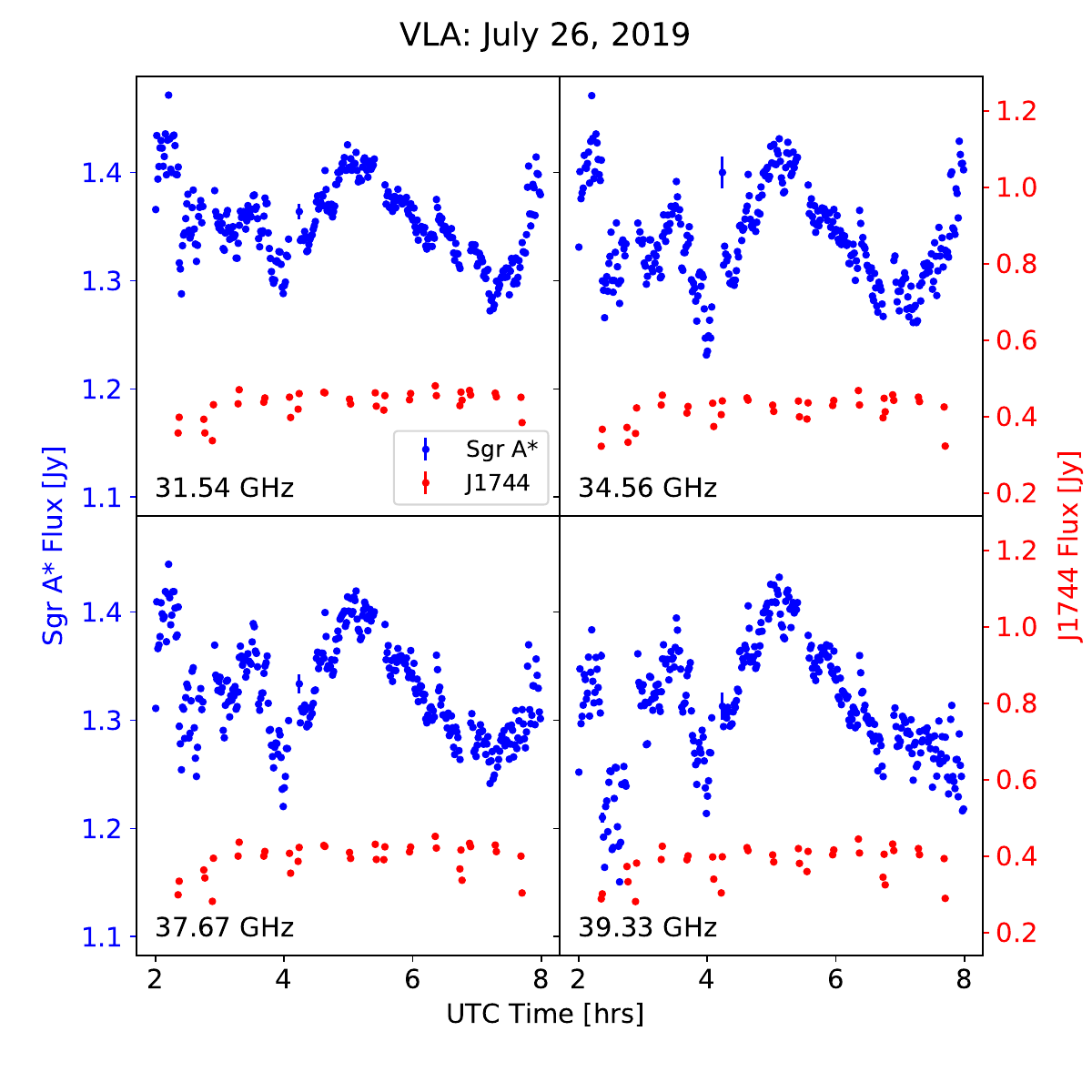}
            \caption{VLA light curves of Sgr A* (in blue) and J1744 (in red) on 26 July 2019 at Ka-band. The error bars are smaller than the points. We bin data at 60-second intervals using projected baselines $\geq100$ k$\lambda$.}
            \label{fig:vla_lcs}
        \end{figure}

        The two nights of observations follow the same procedure with three calibrators and Sgr A*. 3C286 is the flux and polarization angle calibrator. J1733-1304 is the bandpass, secondary complex gain, and primary instrumental polarization calibrator, while J1744-3116 (hitherto denoted J1744) is the primary complex gain and secondary instrumental polarization calibrator. Scans of Sgr A* were interleaved with J1744, and J1733-1304 was observed approximately every hour. The data were reduced in \texttt{CASA} \citep[][v5.6.2]{McMullin2007} with the default VLA pipeline. To calibrate the polarization products, we follow the appropriate polarization calibration guide\footnote{\url{https://casaguides.nrao.edu/index.php?title=CASA_Guides:Polarization_Calibration_based_on_CASA_pipeline_standard_reduction:_The_radio_galaxy_3C75-CASA5.6.2}}. The polarization-calibrated visibilities were exported from \texttt{CASA} and imported into \texttt{AIPS} for additional baseline-based flagging and point-source phase self-calibration. We extract light curves for Sgr A* and J1744 using \texttt{DFTPL} on projected baselines $\geq100~\text{k}\lambda$ at 60-second intervals for each IF. \texttt{DFTPL} averages the real portion of the visibilities at the phase center to determine light curves without the need for other (e.g., imaging-based) techniques. The 26 July 2019 light curves at Ka-band are shown in Figure \ref{fig:vla_lcs}.
        
        \begin{figure*}
            \centering
            \includegraphics[width=\textwidth]{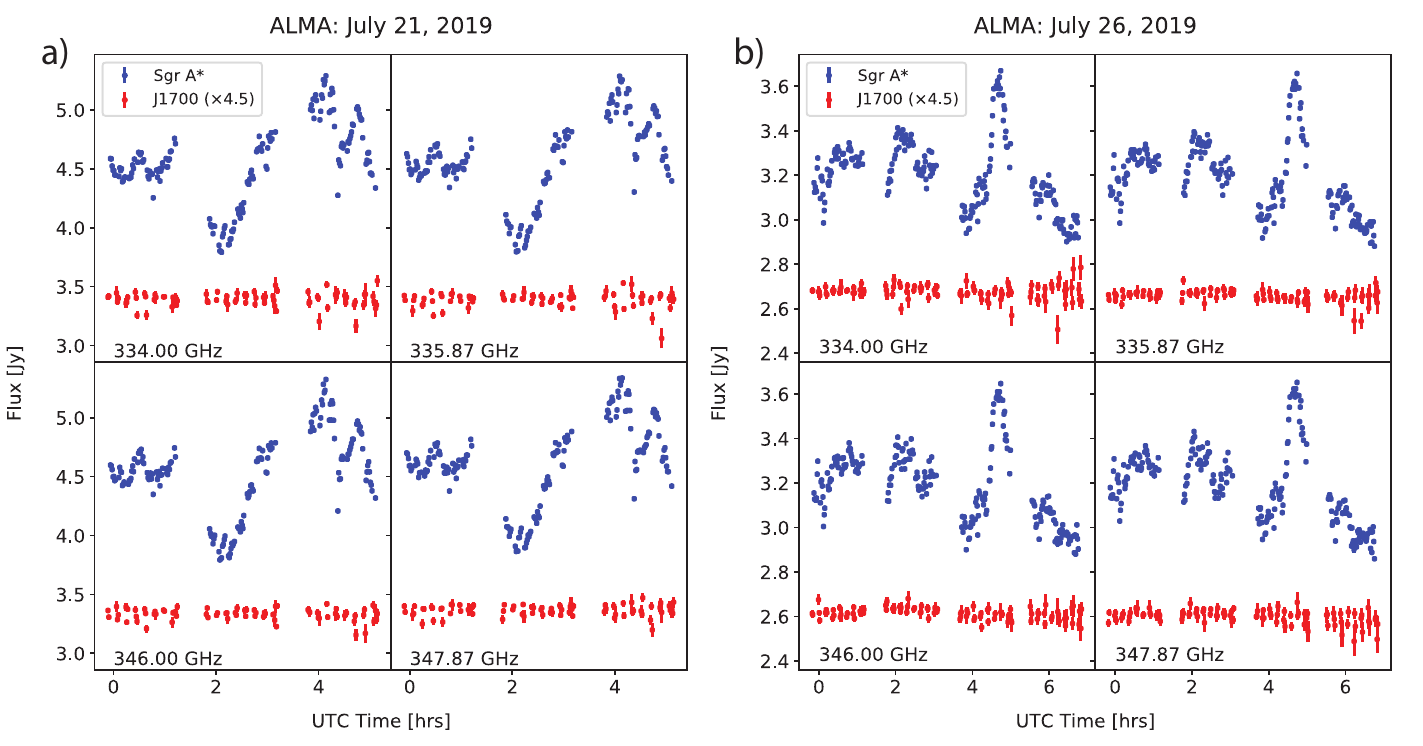}
            \caption{ALMA light curves of Sgr A* (blue) and J1700-2610 (red) for 21 July 2019 (a) and 26 July 2019 (b). All light curves are made with a $60$-second binning time on projected baselines $\geq20$ k$\lambda$.}
            \label{fig:alma_lcs}
        \end{figure*}
    \newpage
    \subsection{Submillimeter: ALMA}
    
        Three nights of ALMA observations were taken as part of the ALMA Compact Array (ACA) Cycle 6 filler program 2018.A.00050.T in band 7 ($\sim330$ GHz) with projected baselines ranging from $10-55~\text{k}\lambda$. We use the \texttt{scriptForPI.py} code to restore the calibrated data from the automatic ALMA pipeline in \texttt{CASA} version 5.4.0-70 (Pipeline-CASA54-P1-B, r42254). Four IFs with 2 GHz of bandwidth each were composed of 1024 1.953-MHz channels recording dual polarization products. One IF is centered on the $\approx346$ GHz CO(3-2, v=0) transition. J1337-1257 and J1924-2914 are the bandpass and amplitude calibrators; J1700-2610 (J1700) and J1717-3342 are the phase calibrators for each observation. Like the VLA data, we phase self-calibrated Sgr A* and J1700, assuming each was a point source to mitigate residual baseline-based phase errors. Finally, we import the data into \texttt{AIPS} and use \texttt{DFTPL} to extract the light curves with a $1$ minute binning time on baselines $\geq20~\text{k}\lambda$. We chose this limit after viewing the visibility amplitude as a function of projected baseline distance, finding that sizes below this threshold are dominated by the extended emission from the Circumnuclear Disk.
        
        We take additional steps to post-process the light curves. We discarded the last execution block on 21 July 2019 due to amplitude-decorrelation on several antennas caused by the low altitude of our calibrators. The calculated flux of the phase calibrator changed (on the order of a few percent) between execution blocks. We follow \citetalias{Michail2021c} to correct this by using the first block as our reference. The phase calibrator's average flux in subsequent execution blocks was used to calculate multiplicative correction factors. These scale factors are applied to the phase calibrator and Sgr A*. We show the final ALMA light curves in Figure \ref{fig:alma_lcs}.

        \begin{figure*}
            \centering
            \includegraphics[width=\textwidth]{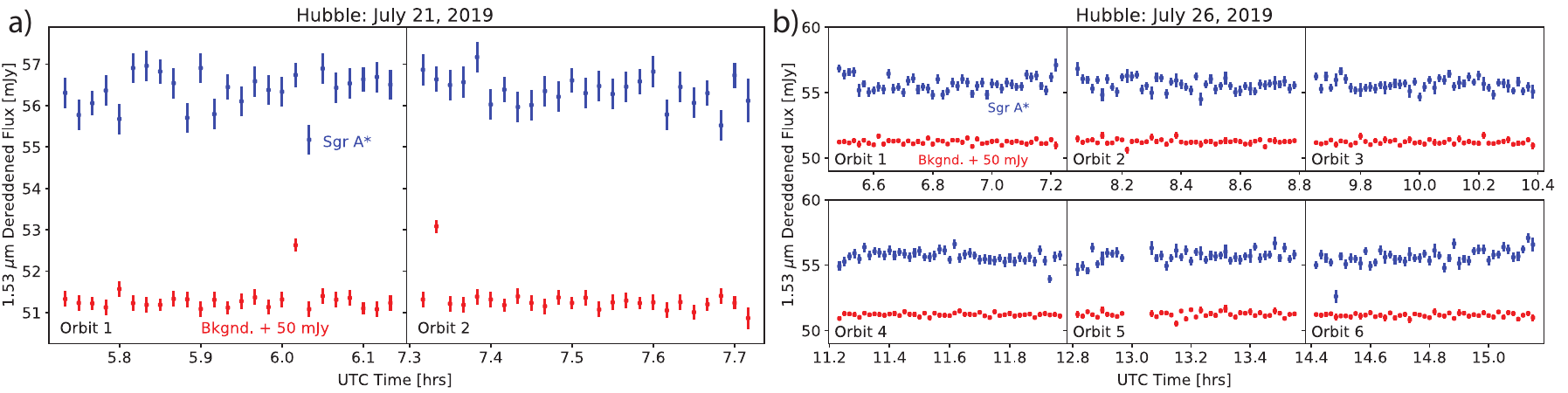}
            \caption{The dereddened HST light curves for Sgr A* (blue) and the background aperture (red) on 21 July 2019 (a) and 26 July 2019 (b) at $1.53$ \micron. Each orbit is separately plotted to show short-term flux variability clearly.}
            \label{fig:hubble_lcs}
        \end{figure*}
        
     \subsection{Near-infrared: Hubble Space Telescope}\label{ssec:hubble}       
        We obtained near-IR observations of Sgr A* with the HST Wide Field Camera 3 (WFC3) under program 15667 (\dataset[10.17909/aqd3-3642]{http://dx.doi.org/10.17909/aqd3-3642}) over two different visits occurring on 21 July (2 orbits) and 26 July (6 orbits). During each visit, Sgr A* was visible to HST at 40--50 minute intervals per orbit. We give a brief explanation of the data calibration here and refer to \citet{Mossoux2016} for further details.

        WFC3 observations were taken with the F153M filter with an effective wavelength of $\lambda_\mathrm{eff}=1.532\ \mu$m taken in continuous accumulation mode. The instantaneous field-of-view is approximately $2\farcm{}26$ per side with a pixel size of $\sim0\farcs{}13$. The accumulation mode reads out the WFC3 detector every 25 seconds during a total exposure time of 275 seconds. HST was dithered around Sgr A*'s location to improve WFC3's point spread function. We calibrate the data with the default STScI \texttt{calwf3} calibration pipeline. We corrected the astrometry of each exposure using the radio location of the nearby IRS-16C source. We used the \texttt{photutils} package of \texttt{astropy} to complete aperture photometry of Sgr A* using a circular aperture of 4 pixels in diameter centered on this source's known radio location at 25-second intervals (the cadence of WFC3's readout mode).

        In our initial analysis, the photometry of other stars in the field in each exposure declined by $\sim3\%$, affecting the final light curve of Sgr A*.  Assuming stars near Sgr A* did not change over a single exposure, we corrected this effect and applied them to Sgr A*'s light curve. We revalidated the light curves for each star and found that they were constant over each exposure. Finally, we applied an aperture correction using several apertures at different sizes by calculating the curve of growth for isolated stars and applying it to Sgr A*'s final light curve.

        Using a reference star, we estimate a noise level of 0.004~mJy per exposure. We deredden the measured fluxes with the prescription in \citet{Fritz2011}. The extinction in the near-IR band is modeled by:
            \begin{equation}
                \text{A}_{\lambda} = 2.62\left(\dfrac{\lambda}{2.166~\micron}\right)^{-2.11}.
            \end{equation}
        For $1.53$ \micron, $\text{A}_{1.53~\mu\text{m}} = 5.46$ mag, giving a flux correction factor of $f_{1.53~\mu\text{m}} = 2.512^{5.46} = 152.8$. We show the dereddened photometry for Sgr A* and the background aperture in Figure \ref{fig:hubble_lcs}. 
        
    \subsection{Mid-infrared: Spitzer Space Telescope}
\begin{figure*}
            \centering
            \includegraphics[width=\textwidth]{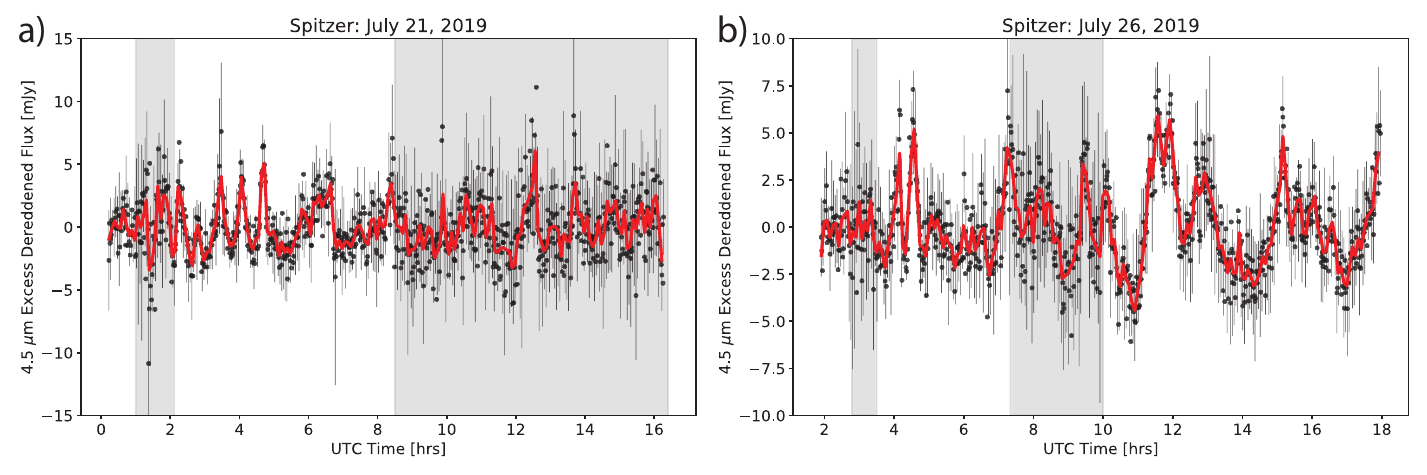}
            \caption{The dereddened \Sp\ light curves at $4.5$ \micron~for Sgr A* during 21 and 26 July 2019 (a and b, respectively). The 1-minute binned data are shown in black, and the 5-minute smoothed data are in red. The gray bands highlight periods of low signal-to-noise fluxes given in \citet{Boyce2022}.}
            \label{fig:spitzer_lcs}
        \end{figure*}

          The \Sp\ observations and calibration were outlined in \citet{Hora2014} and \citet{Witzel2018}, while a more detailed explanation of these data was given in \citet[]{Abuter2021}. Three observations of Sgr A* lasting approximately 16 hours each were taken as part of program ID 14026. In contrast to previous \Sp\ analyses, we bin the data into 1- and 5-minute averages to obtain higher signal-to-noise light curves by averaging all fluxes within one-half bin width. We converted the barycentric UT timestamps into UTC using an online tool\footnote{\url{https://astroutils.astronomy.osu.edu/time/bjd2utc.html}} \citep{Eastman2010}. \citet{Fritz2011} listed a $4.5~\micron$ extinction of $1.06$ mag toward the Galactic Center. Therefore, we scale the \Sp\ light curves by $2.512^{1.06}=2.65$ to deredden the light curves. \citet{Boyce2022} denoted low signal-to-noise-ratio periods in their analysis of these \Sp\ data, which we incorporate here. The light curves for 21 July 2019 and 26 July 2019 are shown in Figure \ref{fig:spitzer_lcs}.
          
    \newpage
    \subsection{X-ray: Chandra X-Ray Observatory}\label{ssec:chandra}
        Three days of simultaneous \Ch\ and \Sp\ observations were completed on 18, 21, and 26 July 2019. For 21 July 2019 and 26 July 2019, the observation IDs are 20446 and 20447 (PI: Fazio), respectively. Each night's observation lasted 52 ks in FAINT mode with Sgr A* centered on the ACIS S3 chip in the 1/8 subarray setting. We reprocessed the data with CIAO version 4.13 and CALDB version 4.9.4. We refine the astrometric accuracy using \texttt{wcs\_match} and \texttt{wcs\_update} with three nearby Galactic Center sources from the Chandra Source Catalog \citep[CSC,][]{Evans2010}.
        
\begin{figure*}
            \centering
            \includegraphics[width=\textwidth]{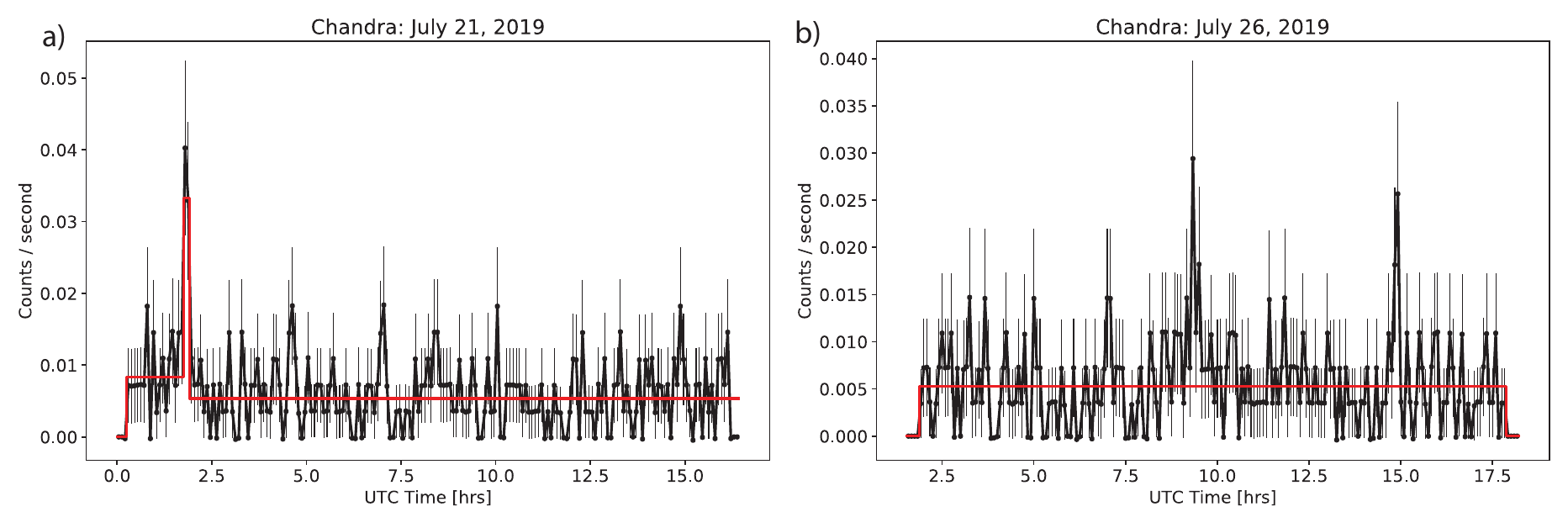}
            \caption{2-8 keV Chandra X-ray Observatory light curves of Sgr A* on 21 (a) and 26 July 2019 (b). The light curves are binned in 300-second increments and are background subtracted. Sgr A* light curves (black) and the Bayesian Block results (red) setting $p_0=0.05$ are shown.}
            \label{fig:chandra_lcs}
        \end{figure*}

        Following \citet{Capellupo2017}, we determine background-subtracted light curves of Sgr A* by placing a $1\farcs25$ radius aperture at the radio location of Sgr A* ($17^\text{h}~45^\text{m}~40\fs04, -29^\circ~00\arcmin~28\farcs18$). Our background aperture is centered on $17^\text{h}~45^\text{m}~40\fs30, -28\degr~59\arcmin~52\farcs66$ with a $10\farcs0$ radius. This background aperture contains no known or obvious sources \citepalias{Michail2021c}. To minimize photons from the diffuse X-ray background, we only use events within the $2-8$ keV range \citep{Neilsen2013}. We used \texttt{dmextract} to obtain the light curves of Sgr A* at a 5-minute binning time and ran the open-source Bayesian block algorithm, \texttt{bblocks}\footnote{\url{https://pwkit.readthedocs.io/en/latest/science/pwkit-bblocks/}}, on the light curves \citep{Scargle1998, Scargle2013, BBlocks} with a false positive rate of $p_0=0.05$. The Bayesian blocks method accounts for counting noise in Poissonian-like systems, where the change in count rate must be statistically significant at a level of $1-p_0$. Since every flare requires two changes \citep[an increase and a decrease;][]{Nowak2012}, this ensures a flare is significant at a confidence of $1-p_0^2 = 99.75\%$.  We show the \Ch\ light curves and the Bayesian block results in Figure \ref{fig:chandra_lcs}.
        
\section{Analysis}\label{sec:analysis}   
    \subsection{21 July 2019}\label{ssec:july21}
        \begin{figure*}
            \centering
            \includegraphics[width=\textwidth]{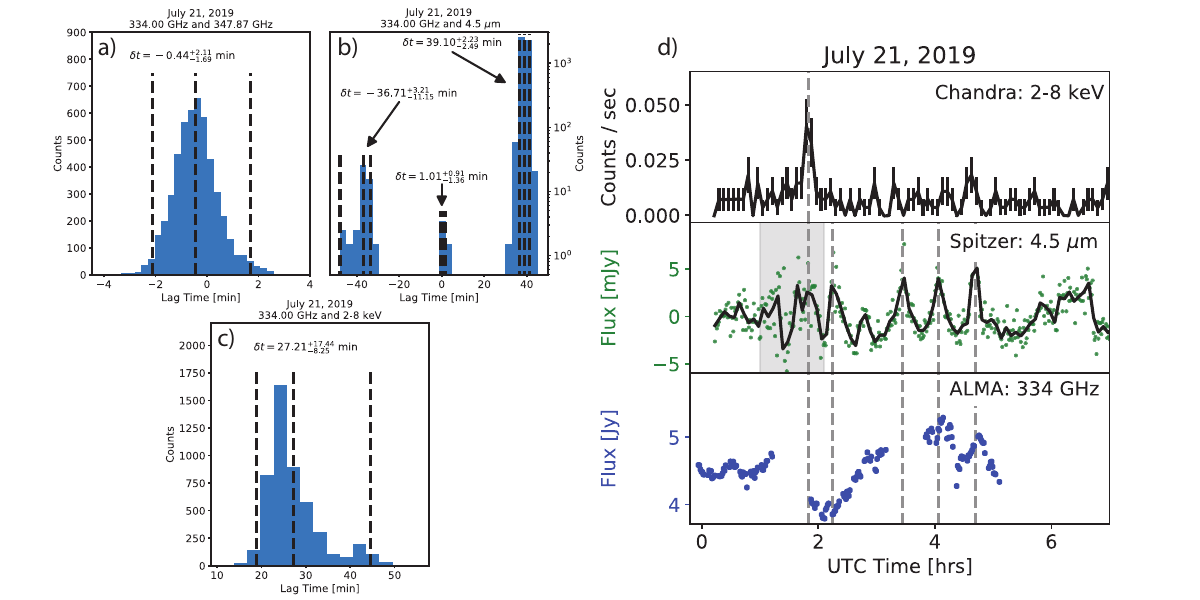}
            \caption{Cross correlations for each pair of observations on 21 July 2019. (a): The cross correlation histogram across the ALMA bandwidth. (b): The cross correlation histogram between the lowest ALMA frequency and \Sp. (c): The cross correlation histogram between the (inverted) ALMA and X-ray light curves. (d): Stacked X-ray, IR, and submm light curves during ALMA coverage. Vertical, gray dashed lines indicate the timestamps of peaks in the X-ray/IR light curves. In the IR plot, the green dots and solid black line represent the 1-minute and smoothed 5-minute light curves, respectively.}
            \label{fig:july21_xcorr}
        \end{figure*}

        Sgr A* was remarkably active on 21 July 2019 in the submm, IR, and X-ray. The Bayesian blocks algorithm (Figure \ref{fig:chandra_lcs}a) detected an increased quiescent plateau and flare in the X-ray light curves, which then decays to a count rate consistent with \citet[][$\sim5$ cts/ks]{Neilsen2015}. 
        To detect flaring events in the HST data, we use a root-mean-square-based (RMS-based) calculation. For each orbit, we calculate the RMS flux of Sgr A* and the background RMS error (RMSE) to quantify the intrinsic scatter in the data. Then, we determine the z-score for each associated flux and error measurement using Sgr A*'s average flux and the background RMSE in the given orbit. We define a flare-like event when the z-score is $\geq5\sigma$ (the typical HST RMSE per orbit is $\sigma\sim0.2$ mJy). For 21 July 2019, we do not detect any flares in the $1.53$ \micron\ data. Therefore, we only focus on the ALMA, \Sp, and \Ch\ light curves.
        
        In Figure \ref{fig:july21_xcorr}, we show cross-correlations histograms between different pairs of the data (panels a-c) and a subset of these three light curves during the ALMA observations (panel d). We detect several corresponding features that occur in both the \Sp\ and ALMA light curves following a single X-ray flare. The lack of \Sp\ data at the peak of the X-ray emission is due to low-SNR data (gray band), so we are unable to definitively conclude if there is an IR flare. Multiwavelength campaigns have shown X-ray flares always appear paired to an IR peak \citep[][]{DE2010}. Therefore, we assume a weak IR flare occurs but is not detected.

            \subsubsection{Cross Correlations}\label{sssec:21jul_cc}
                We complete cross-correlation analyses to check for time-delayed emission using \texttt{pyCCF} \citep{Peterson1998, PYCCF}. For all cross-correlations in this paper, we use 5000 Monte Carlo realizations that remove a random number of light curve points each time and build a time delay histogram between two light curves. We report the mean delay value and the $95\%$ ($2\sigma$) confidence interval. We consider that the light curves are delayed if zero is not contained within the 95\% interval. 
        
                We show the three cross-correlation histograms for 21 July in Figure \ref{fig:july21_xcorr}a-c. We report no statistically significant time delay across the ALMA bandwidth ($\sim14$ GHz). \citet{Iwata2020} and \citet{Wielgus2022a} suggest the presence of a small (but detectable) time delay across their lower frequency 230 GHz bandwidth. Using Equation 2 and the values listed in \citet{Iwata2020}, we calculate an expected time delay across the 345 GHz ALMA band on the order of $\sim30$ seconds, assuming a range of electron indices ($p$) between $1-3$. Since we average the light curve in 60-second increments, any time delays shorter than this will be washed out.
                
                Cross correlations are sensitive to peaks in emission (i.e., when the correlation coefficient is near $1$) instead of a peak and trough (i.e., when the correlation coefficient is near $-1$). We invert the ALMA light curve before completing the cross-correlation between the submm and X-ray light curves to determine the time delay between the X-ray flare peak and the $\sim0.7$ Jy drop in submm flux. We find the X-ray data lead the submm light curves by approximately $26$ minutes. We mark the X-ray peak flare time with a black dashed line in Figure \ref{fig:july21_xcorr}d to clearly show this detected delay. Such a feature has been detected previously several times \citep[i.e.,][]{Marrone2008, Meyer2008, FYZ2008, Capellupo2017} and will be further discussed in Section \ref{ssec:occultation}.
        
                Finally, we complete the cross-correlation between the ALMA and \Sp\ light curves (Figure \ref{fig:july21_xcorr}b); the time delay histogram shows a much different structure than in the other two pairs of light curves. We find three groups of time delay peaks (centered near $\pm35$ and $0$ minutes). The majority of the 5000 realizations prefer the positive time delay where the IR leads the submm. The negative time delay peak (where the submm leads the IR) appears because \texttt{pyCCF} is model-agnostic and corresponds to the third IR peak corresponding to the first observed one in the submm. For our analysis, it has no physical meaning.
        
\begin{figure*}[th]
                    \centering
                    \includegraphics[trim= 1.cm 0.25cm 0.75cm 1.25cm, clip, width=\textwidth]{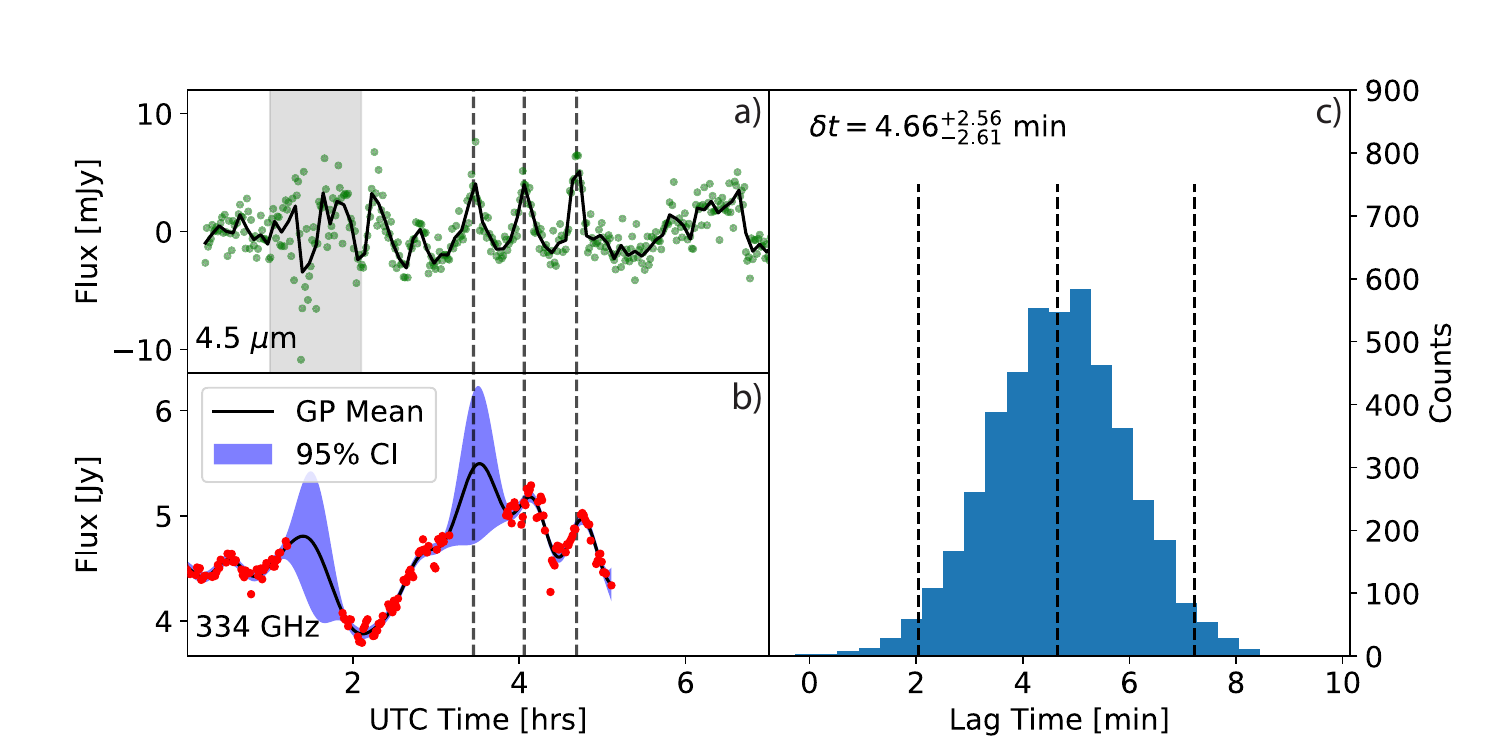}
                    \caption{Submm and IR light curves of Sgr A* on 21 July 2019. (a): \Sp\ light curve during the significant period of variability present in both the IR and submm data. The 1-minute and smoothed 5-minute light curves are shown in green and black, respectively. (b): The raw (red) and mean interpolated (black) ALMA light curves during the same time range as above. We interpolated the raw light curves using a Gaussian process, and the 95\% confidence interval (CI) is shaded in purple. (c): The \texttt{pyCCF} lag time histogram between the \Sp\, and interpolated ALMA data. We find the submm data are delayed $\delta t=4.66^{+2.56}_{-2.61}$ minutes relative to the IR data.}
                    \label{fig:alma_spitzer_gp}
                \end{figure*}

                Due to the lack of ALMA data near 3.5 hr UTC, we cannot rule out the possibility of simultaneity. To determine if peaks in the submm and IR are simultaneous, we follow \citetalias{Michail2021c} in using a Gaussian process regression to interpolate the ALMA data between execution blocks. We use the Gaussian process toolkit, part of the \texttt{sklearn} Python package, and choose a compound constant + radial-basis kernel. In Figure \ref{fig:alma_spitzer_gp}a,b, we show the Gaussian process-regressed ALMA and \Sp\ data. The purple band denotes the 95\% confidence interval of the Gaussian process fit, and the black marks the mean fit. The red points are the original (non-interpolated) data. We again use \texttt{pyCCF} to search for time lags between the interpolated ALMA and \Sp\, data; the cross-correlation histogram is shown in Figure \ref{fig:alma_spitzer_gp}c. We utilize the 68\% ($1\sigma$) confidence interval as the interpolated data error inputs to \texttt{pyCCF} instead of the 95\% interval for consistency with Figure \ref{fig:july21_xcorr}. From this interpolated analysis, we find the IR data slightly lead the submm peaks by $4.66^{+2.56}_{-2.61}$ minutes and suggests the presence of an ALMA flare near 3.5 hr UTC occurring between execution blocks.
        
        \subsubsection{Periodograms}     
        
            \begin{figure*}
                \centering
                \includegraphics[width=\textwidth]{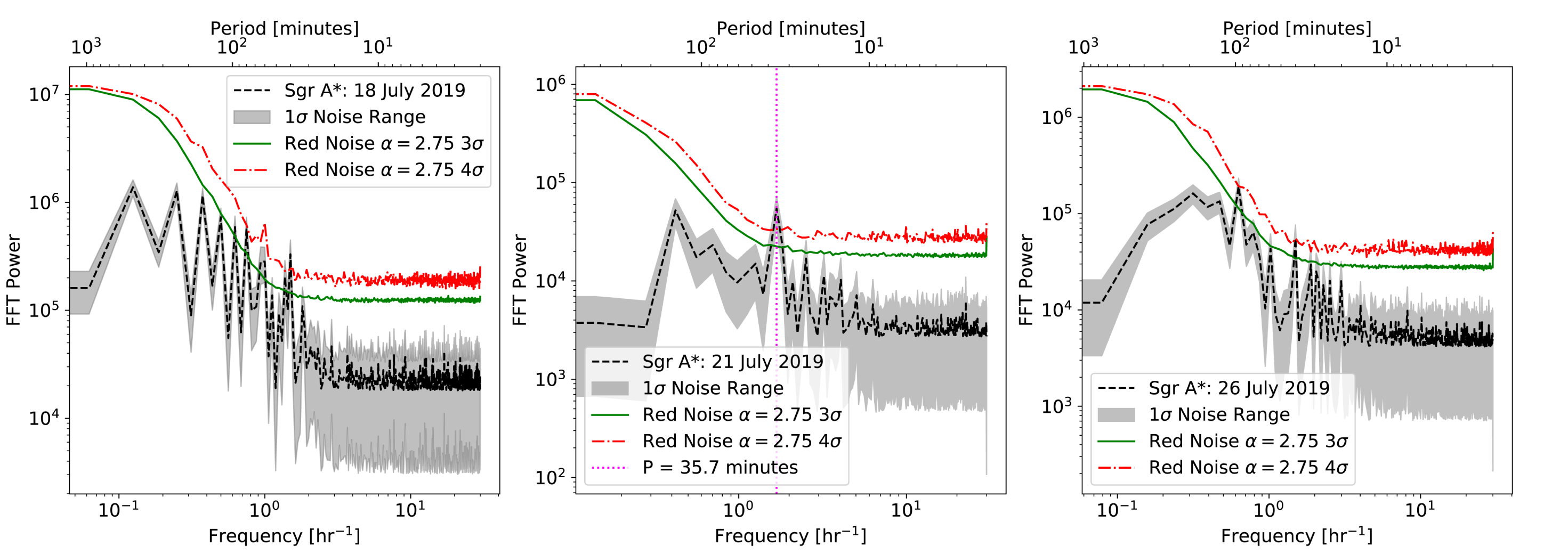}
                \caption{Monte Carlo Fourier transform power spectra for 35,000 realizations of Sgr A* and red noise during all three days of observation. The black dashed line is the mean power spectra from the Sgr A* light curve, while the gray shaded region shows the 68\% confidence interval. $3\sigma$ (green solid) and $4\sigma$ (red dot-dash) upper-limit power spectra are shown for red-noise light curves with the same statistics as Sgr A*. We detect a transient periodicity at $>4\sigma$ significance at $\sim35$ minutes on 21 July 2019 and a slightly significant $\sim40$ minute peak on 18 July 2019. The dotted magenta line in the middle panel denotes the statistical peak of the periodogram.}
                \label{fig:alldata_lomb_scargles}
            \end{figure*}
            
            We now focus on the repeating flares present in the mid-IR and submm light curves. The initial cross-correlation results (Figure \ref{fig:july21_xcorr}) between these two wavelengths suggest periodic emission occurring every $\approx40$ minutes. The \Sp\ data were binned on equal intervals of 1 minute, so we use \texttt{numpy} fast Fourier transforms (FFT) to detect evidence for transient periodicity. However, these algorithms have neither error-weighting or significance-testing built in. We discuss the approach to each of these topics below.

            To account for the photometric scatter in \Sp's light curve, we complete a Monte Carlo process to ``error weight'' the power spectra. To do this, we randomly resample each of the observed points within their (normally distributed) flux uncertainties and redo the FFT on the ``new'' light curve. Completing this process 35,000 times allows us to probe a probability of $\sim1/35000 \approx4\sigma$; we report the mean and $1\sigma$ CI power spectra as the error-weighted power spectrum.

            To determine if any peaks in the power spectrum are significant, we follow the prescription in \citet{Do2009} to compare the measured FFT to a red-noise-dominated power spectrum. We generate 35,000 red-noise light curves (using the \texttt{stochastic}\footnote{\url{https://stochastic.readthedocs.io/en/stable/index.html}} Python package) with a spectral density exponent $\alpha=2.75$ \citep[PSD $\propto f^{-2.75}$;][]{Weldon2023}. These red-noise light curves have the same temporal sampling as our light curves. We scale each of the generated light curves to have the same mean and standard deviation as the Sgr A* light curve. To account for measurement noise in the light curve, we re-sample each ``flux'' point with the corresponding uncertainty in the \Sp\ data. Finally, we calculate the power spectra for each generated light curve and report the $3\sigma$ and $4\sigma$ upper limits on the power. We consider a statistically significant detection in the periodogram when Sgr A*'s mean spectral power is greater than the $4\sigma$ red-noise spectrum. In Figure \ref{fig:alldata_lomb_scargles}, we show the power spectra for Sgr A* during all three days using all data (outside of the low-SNR ranges) compared to the generated red noise light curves.

            On 21 July 2019, we find a significant peak at a $\sim35$ minute period with a similar (but weaker) peak at $\sim40$ minutes on 18 July 2019. These timescales are similar to \Sp's quasi-periodic pointing wobble with a period of $\sim40$ minutes, which would cause photometric amplitude variations of a couple of percent for a point source \citep[see, e.g.,][for more complete descriptions of the wobble]{Grillmair2012, Grillmair2014}. This effect was present for earlier observations of Sgr A* but was largely corrected, mitigating the resulting photometric response. Figures 8 and 2 in \citet{Hora2014} show the photometric changes caused by the wobble and the light curve of Sgr A* after correction, respectively.
            
            There are several key reasons why we trust the validity of this transient periodicity. First, the statistically significant signal only occurs in one of the three days, specifically on the same observation where an identical peaked structure is observed in the ALMA light curves. Second, a more comprehensive catalog of \Sp\ light curves was published in \citet{Witzel2021}, where no quasi-periodicity was detected. A wobble with a sustained period would have been detected in their analysis at higher significance since they used \Sp\ data beginning in late 2013. Additionally, near the end of \Sp's mission, the heating duty cycle was changed because it was in a different thermal state where temperature cycling did not occur, rendering this pointing wobble absent. Finally, in the absence the FFT results, the cross correlation plots between ALMA and \Sp\ (Figure \ref{fig:july21_xcorr}b) and their light curves (Figures \ref{fig:july21_xcorr}d and \ref{fig:alma_spitzer_gp}a,b) still show a similar $35$ minute timescale. Therefore, the sequential peaks observed in \Sp\ on the 21st may be suggestive of several explanations, such as footprints of flux tubes within the accretion flow \citep[i.e.,][]{Porth2021} or short-lived, orbiting, transient hotspot within the accretion flow. We discuss the latter model further in Section \ref{ssec:trans_period}.
        
    \subsection{26 July 2019}\label{ssec:26jul}
        \begin{figure*}
            \centering
            \includegraphics[width=\textwidth]{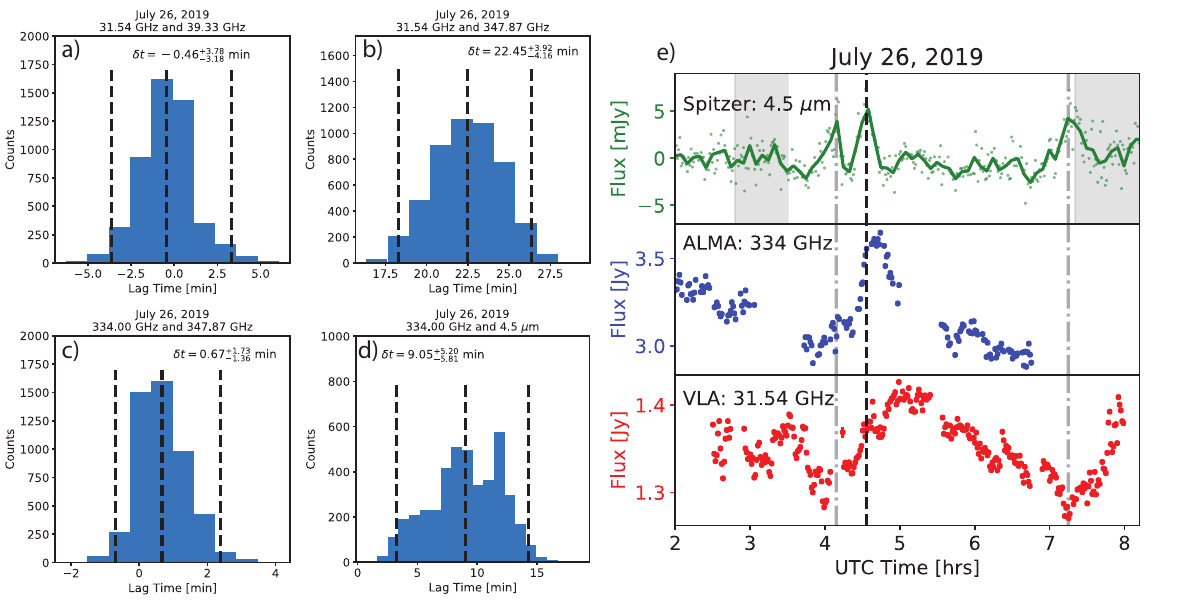}
            \caption{Same as Figure \ref{fig:july21_xcorr}, but for the 26 July 2019 observations. (a): Cross correlation histogram across the VLA bandwidth. (b): Cross correlation between the most widely separated VLA and ALMA frequencies. (c): Cross correlation across the ALMA bandwidth. (d): Cross correlation between the ALMA and \Sp\ light curves. (e): Stacked \Sp\,, ALMA, and VLA light curves centered on the dominant observed flare. The vertical black dashed line corresponds to the peak IR flux. The gray dash-dotted lines denote times of IR flares where there are possible dips in the radio light curve; we do not study these unconfirmed events here. The 1-minute and smoothed 5-minute \Sp\ data are displayed by dots and the solid green line, respectively.}
            \label{fig:july26_xcorr}
        \end{figure*}
        
        We use the same RMSE-based analysis in Section \ref{ssec:july21} to detect flares in the HST data on this day (Figure \ref{fig:hubble_lcs}b). Aside from two points ($t\approx 11.95, 14.5$ UTC hrs), the variations in the 1.53 \micron\ light curves are within $5\sigma$ RMSE of the mean flux during each orbit. These data points are below the mean and do not correspond to variations in the \Sp\ light curves. Therefore, we find no evidence for flaring events at $1.53$ \micron. These non-detections allow us to place an upper limit on the IR flare spectral index using \Sp\ and HST. Orbits 5 and 6 on this day (Figure \ref{fig:hubble_lcs}b) overlap with moderately weak \Sp\ flares of $\sim5$ mJy at 4.5 \micron. The upper limits of $1$ mJy from HST (i.e., $\sim5\times$RMSE per orbit) imply a spectral index of $\alpha\lesssim-1.5$ between 1.53 and 4.5 \micron. This value is steeper than the canonical $\alpha=-0.7$ found between 1.6 and 2.2 \micron\ \citep{Bremer2011}, but comparable to the spectral index between 1.6 and 4.5 \micron\ at the beginning of the bright flare observed on 18 July 2019 \citep[$\alpha\sim-1.45$;][]{Abuter2021}.
        
        We detect no X-ray flaring events in the 2-8 keV energy range from the Bayesian blocks analysis (Figure \ref{fig:chandra_lcs}b). To check for lower significance flares, we test values of $p_0\in\left[0.05, 0.95\right]$ in increments of $0.05$; however, none were detected. This places an upper limit of $1.45$ flares day$^{-1}$ and matches recent published ranges \citep[$\sim1.1-1.6$ flares day$^{-1}$;][]{Neilsen2013, Ponti2015, Yuan2016, Mossoux2017, Bouffard2019}.
        
        The \Sp\ light curves show a great deal of variability, particularly during the latter half of the observation (Figure \ref{fig:spitzer_lcs}b). However, since the radio and submm light curves occur during the first half, we do not focus on these later variations. A flare event in the radio and submm is clearly present (Figures \ref{fig:vla_lcs} and \ref{fig:alma_lcs}b, respectively). 

        \subsubsection{Cross Correlations}
            We again use \texttt{pyCCF} to check for time-delayed emission in four different frequency pairs: radio-radio, radio-submm, submm-submm, and submm-mid-IR. The histograms for the cross-correlations are shown in Figure \ref{fig:july26_xcorr}a-d. We do not find evidence for a time delay across the radio and submm frequency bandwidths. However, the lowest frequency radio emission lags $\delta t=22.45^{+3.92}_{-4.16}$ minutes behind the highest frequency submm light curve. We further find the submm peak delayed $\delta t=9.02^{+4.98}_{-5.82}$ minutes behind the mid-IR peak. In Figure \ref{fig:july26_xcorr}e, we present the stacked mid-IR, submm, and radio light curves, which clearly show the time delay evolution of the flare.

            We examined the lack of time delay across the ALMA bandwidth in Section \ref{sssec:21jul_cc}. However, using the values in \citet{Iwata2020} would suggest a $\sim9$ minute time delay across the VLA bandwidth, which we do not detect (Figure \ref{fig:july26_xcorr}a), but should be measurable. Time delays across a single VLA band have been previously detected, such as the 20-minute lag between 8 and 10 GHz found by \citet{Michail2021b}. Recalculating the expected time delay using Equation 2 from \citet{Iwata2020} with the measured 22.5-minute lag between 31.45 GHz and 347.87 GHz (Figure \ref{fig:july26_xcorr}b) yields a theoretical $\sim3$ minute delay across the VLA bandwidth (assuming $p$ ranges from $1-3$). We argue that the absent radio time delay is caused by observational effects in the VLA data rather than a breakdown in the adiabatically-expanding hotspot picture. In Section \ref{ssec:vla}, we note the numerous weather- and setup-related issues that plagued the VLA observations; these will tend to increase the scatter (noise) in the light curves. Since a 3-minute time delay relies on only three points per light curve (60-second averaging), this, compounded with the increased scatter, restricts our ability to detect delayed emission. The $2\sigma$ scatter of the VLA's cross correlation histogram (Figure \ref{fig:july26_xcorr}a) matches the expected 3-minute delay, implying the lag may be hidden in the noise.

        \subsubsection{Additional Light Curve Features}
            We note the presence of two additional features (marked with gray dash-dotted lines in Figure \ref{fig:july26_xcorr}e) in the multiwavelength light curves aside from the prominent flare in Figure \ref{fig:july26_xcorr}. Near $4.1$ hr UTC, an IR flare occurs, coincident with a slight ($\sim100$ mJy) increase in the submm flux but a drop ($\sim50$ mJy) in the radio flux. It is unclear if the radio feature indicates an occultation-like event (where the flare hides a portion of the accretion flow as seen from the Earth) or an issue with the amplitude scaling between observing blocks.  
            
            A tertiary feature appears in the radio light curve near $7.25$ hr UTC, coincident with an increase in the IR. It is unclear if this is another occultation-like event or the beginning of an additional flare. The lack of submm coverage, the proximity of the IR feature to the beginning of a low-SNR part of the light curve (gray band), and not observing the radio peak all blur its final classification. Because of these issues, we do not further consider this feature in our discussion.

\section{Models of Observed Events in the Light Curves}\label{sec:models}
    In this Section, we aim to model the submm flux decrease event observed on 21 July 2019 (Figure \ref{fig:july21_xcorr}) and the time-delayed flare detected in the IR, submm, and radio regimes on 26 July 2019 (Figure \ref{fig:july26_xcorr}). While several interpretations aim to describe the anti-correlation between submm and IR/X-ray light curves, we will focus on the occultation model. Throughout the descriptions below, we assume that Sgr A* is comprised of two components, namely the variable and background (quiescent) emission. The variable component generates the flares, while the quiescent component is responsible for the non-zero flux offset and low-amplitude, long-timescale variability.

    \begin{figure}
        \centering
        \includegraphics[width=0.45\textwidth]{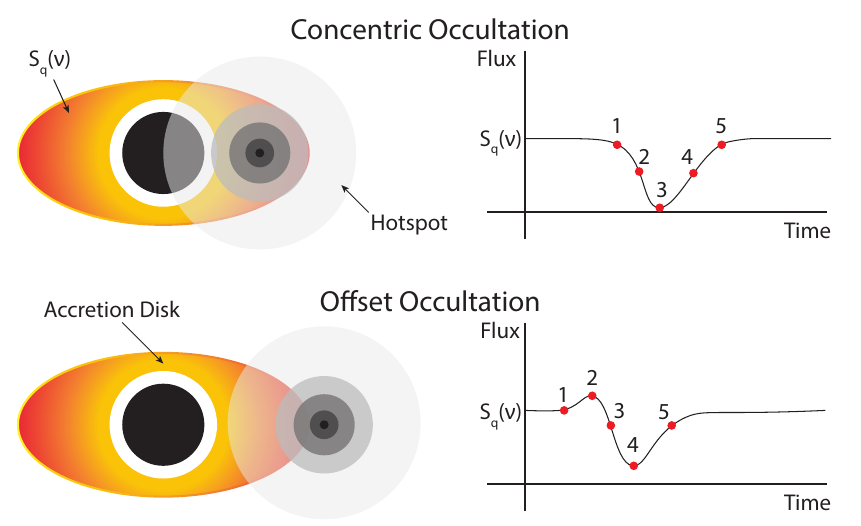}
        \caption{The two possible setups for an occultation to occur. The accretion flow (orange) is shown surrounding Sgr A* (black circle). $S_q(\nu)$ denotes the accretion disk flux at frequency $\nu$. The size of the adiabatically-expanding hotspot at several times is shown in gray. The resulting light curves, as seen by an observer who is face-on to the page, are shown in the diagrams on the right. \textit{Top}: A concentric occultation occurs when the hotspot is fully embedded within the accretion flow. \textit{Bottom}: An offset occultation is shown, where the hotspot does not begin to cover the accretion flow until some time later. Both a flare peak (number 2) and dip (number 4) occur.}
        \label{fig:occultation_diagram}
    \end{figure}

    \subsection{Occultation of the Accretion Flow by Adiabatically-Expanding Synchrotron Hotspots}\label{ssec:occultation_model}
        In Figure \ref{fig:july21_xcorr}, we document another instance of decreasing submm flux during an IR or X-ray flare. \citet{FYZ2010} models this anti-correlation as a geometric occultation of Sgr A*'s accretion disk by a flare that is optically thick in the submm. The flaring region is in front of the accretion disk seen from the Earth, eclipsing a fraction of its projected area on the sky, causing a dip in the submm light curve. The general setup for this model is depicted in the top panel of Figure \ref{fig:occultation_diagram}. The initial description of the model only accounted for a specific electron power-law index; here, we generalize their model for all values.

        \subsubsection{Hotspot Equations}
            We assume the flare is produced by self-absorbed synchrotron emission from an electron spectrum $n(E)~dE\propto E^{-p}~dE$. We further assume the hotspot is a homogeneous sphere with an initial radius $R$ and contains an initial magnetic field strength $B$. For a power-law electron spectrum, the synchrotron source function, $J_\nu$, is
                \begin{align}
                    J_\nu = \dfrac{\sqrt{2}}{(p+1)} \frac{\Gamma\left(\frac{3p-1}{12}\right)\Gamma\left(\frac{3p+19}{12}\right)}{\Gamma\left(\frac{3p+2}{12}\right)\Gamma\left(\frac{3p+22}{12}\right)}\dfrac{\nu^2}{c^2}E_{\nu},
                    \label{eq:J_nu}
                \end{align}
            where $E_\nu$ is the characteristic electron energy
                \begin{align}
                    E_\nu = \left(\dfrac{4\pi m_e c\nu}{3Bq}\right)^{1/2}m_ec^2.
                    \label{eq:e_nu}
                \end{align}
            $m_e$, $c$, and $q$ represent the electron mass, speed of light, and (absolute) electron charge, respectively. The hotspot of radius $R$ spans a solid angle $\Omega = \pi\left(R/d\right)^2$, where $d$ is the distance to Sgr A*.
            
            We define $\tau_\nu$, the frequency-dependent optical depth, as $\tau_\nu = (4/3)\kappa_\nu R$. The $4/3$ prefactor accounts for the ``effective'' optical depth of all lines of sight through the plasma. This is different than the canonical $\tau_\nu = 2\kappa_\nu R$, which is only true for the optical depth through the plasma's center. $\kappa_\nu$ is the synchrotron absorption coefficient,
                \begin{align}
                    \kappa_\nu = \dfrac{\sqrt{3}Bq^3}{4\pi m_e\nu^2}\left(2^{p/2-1}\Gamma\left(\dfrac{3p+2}{12}\right)\Gamma\left(\dfrac{3p+22}{12}\right)\right)~n(E).
                    \label{eq:kappa_nu}
                \end{align}
            We have set the pitch angles to $90^\circ$ in Equations \ref{eq:e_nu} and \ref{eq:kappa_nu}. To adopt a fixed ratio between the magnetic field energy density ($U_B = B^2/8\pi$) and the electron energy density ($U_e = \int_{E_1}^{E_2} En(E)~dE$), we define the electron number density between initial energies $E_1$ and $E_2$ as
                \begin{align}
                    n(E) = \dfrac{B^2}{8\pi\varphi E_0^2}\left(\dfrac{E}{E_0}\right)^{-p}.
                    \label{eq_nE}
                \end{align}
            $\varphi$ is the equipartition factor and controls the ratio of the energy densities between the magnetic and particle fields. We derive $\varphi$ and the prefactor for Equation \ref{eq_nE} in Appendix \ref{appendix:a}. As the occulting plasma expands, the power-law bounds scale inversely with radius \citep[i.e., $E_1' = E_1(R / R_0)^{-1}$,][]{VDL1966}, caused by adiabatic cooling of the ultra-relativistic electron gas.  For this analysis, we assume $E_0 = E_1 = 1$ MeV and $E_2 = 1$ GeV. We note the special constraint for the energy bounds where $E \nless 0.5$ MeV (electron rest mass), which we account for in our analysis. 
            
            The measured flux of the hotspot at any radius and frequency:
                \begin{equation}
                    S_\nu(R) = \Omega J_\nu\left(1-e^{-\tau_\nu}\right).
                \end{equation}
            To convert between radius and UTC time, we assume a linear expansion of the hotspot with expansion speed $v_{\text{exp}}$:
                \begin{equation}
                    \dfrac{R}{R_0} = 1 + v_{\text{exp}}(t - t_0),
                    \label{eq:radius_time_relationship}
                \end{equation}
            where $t_0$ is the UTC time when $R=R_0$.
            
        \subsubsection{Quiescent Emission Model}
            In this geometric model, the fraction of the projected accretion flow's area occulted by the hotspot is denoted $f$. From previous analyses, the intrinsic area of the accretion flow is modeled by a frequency-dependent power-law, which we define as $A_{q}(\nu)$. We use the results of \citet{Cho2022} to find $A_q(\nu) = 6.1(\nu/300\text{~GHz})^{-2.4}~R_{\text{S}}^2$ assuming a Gaussian profile having area $A_q(\nu) = \left(2\pi/8\ln2\right) \theta_{\text{major}}(\nu)\theta_{\text{minor}}(\nu)$. The choice of $A_q$ will determine if there is an explicit inclination term. The \citet{Cho2022} analysis gave the intrinsic size of Sgr A* projected on the sky, removing the need for a specific inclination term. Placing the hotspot at the center of the accretion flow lets us define the ratio $f \equiv \pi R^2/A_q$ if $R\leq \sqrt{A_q/\pi}$ and $f=1$ otherwise, measuring the fraction of the accretion flow eclipsed by the hotspot. We note \citet{Cho2022} use only one epoch of data to determine the frequency-dependent size of Sgr A*. \citet{Cheng2023} observed Sgr A* over 10 months, finding a variable intrinsic area of $\approx13\%$, while the mean was roughly $10-20\%$ higher than the \citet{Cho2022} single-epoch results. However, this difference would not substantially affect our results.

            In addition to the intrinsic size, the accretion flow has a characteristic flux at frequency $\nu$ denoted by $S_{q}(\nu)$. This model could be further generalized to include a time-dependent term, but that is outside the scope of this analysis. We assume the accretion flow is uniformly smooth across the entire projected area.
            
        \subsubsection{Total Intensity Light Curve}
            There are three sources of flux present in this model: the unobscured and obscured accretion flow and the hotspot. The unobscured flow has flux $\left(1-f\right)S_q$ and the hotspot $\Omega J_\nu\left(1-e^{-\tau_\nu}\right)$. The obscured flow flux passes through the optical depth produced from the hotspot yielding a flux $fS_q e^{-\tau_\nu}$. Altogether, the net sum of these terms yields a total observed flux of
                \begin{align}
                    S(\nu) &= S_{q}(\nu) + \left(\Omega J_\nu - fS_{q}(\nu)\right)\left(1-e^{-\tau_\nu}\right).
                    \label{eq:total}
                \end{align}
            We note the parameters $B, \Omega, J_\nu, f, \tau_\nu, \varphi, E_1, \rm{ and } ~E_2$ all vary with radius, and, therefore, time (Equation \ref{eq:radius_time_relationship}). When modeling this time-dependent nature of the occultation, the secular changes in these parameters must be accounted for.
    \\
    \subsection{Flaring Emission Produced by Adiabatically-Expanding Synchrotron Hotspots}\label{ssec:flaring_model}
        We model the multiwavelength flare using an adiabatically-expanding synchrotron hotspot as described by \citet{VDL1966} and \citet{FYZ2006}. To characterize both the quiescent and variable emission, we use a joint fit of accounting for secular variations in the accretion flow and the time-dependent nature of the flaring emission.

        \subsubsection{Flaring Emission Model}
            The variable model is characterized by four parameters: $S_0$, $t_0$, $p$, and $v_{\rm{exp}}$. $S_0$ is the peak flux density at frequency $\nu_0$, which is produced by an electron spectrum $n(E)~dE\propto E^{-p}~dE$ between energies $E_{\rm{min}}$ and $E_{\rm{max}}$. The flaring region expands at a (normalized) velocity $v_{\rm{exp}}$. To convert between UTC time and normalized radius, we use Equation \ref{eq:radius_time_relationship}. 
                
            The flux density of the flare at any radius and frequency $\nu$ is given by
                \begin{align}
                    S_\nu(t) = S_0 \left(\dfrac{\nu}{\nu_0}\right)^{2.5}\left(\dfrac{R}{R_0}\right)^3\dfrac{1 - e^{-\tau_\nu}}{1-e^{-\tau_0}},
                \end{align}
            where $\tau_\nu$ and $\tau_0$ are the frequency/radius-dependent and critical optical depths, respectively. For a given $p$, the critical optical depth can be determined from the following transcendental relationship:
                \begin{align}
                    e^{\tau_0} - \left(\dfrac{2p}{3}+1\right)\tau_0 - 1 = 0.
                \end{align}
            $\tau_\nu$ at any frequency and radius is determined via
                \begin{align}
                    \tau_\nu = \tau_0 \left(\dfrac{R}{R_0}\right)^{-(2p+3)} \left(\dfrac{\nu}{\nu_0}\right)^{-(p+4)/2}.
                    \label{eq:tau_nu}
                \end{align}

        \subsubsection{Quiescent Emission Model}
            To model time-dependent secular and frequency-dependent variations in the accretion flow, we use a linear model of the form
                \begin{align}
                    I_q(\nu, t) = \left(I_0^\nu + \delta I_t^\nu \left(t - t_0\right)\right)\left(\dfrac{\nu}{\nu_0}\right)^{\alpha_{\nu}}.
                    \label{eq:quiescent}
                \end{align}
            Here, $I_0^{\nu}$ is the quiescent emission flux density at frequency $\nu_0$ at time $t_0$, $\delta I_t^\nu$ is the time-slope of the quiescent flux, and $\alpha_{\nu}$ is the quiescent emission's spectral index. 
    
\section{Discussion}\label{sec:discussion}

    In this Section, we consider the periodic signal detected in the IR and hinted at in the submm in the context of previous analyses. Using the frameworks described in Section \ref{sec:models}, we then model the occultation event and the detected multiwavelength flare. Finally, we discuss a caveat in our modeling and quantify how it affects our results.
    
    \subsection{21 July 2019}
        \subsubsection{Transient Periodicity}\label{ssec:trans_period}
        
            \begin{figure}
                \centering
                \includegraphics[trim=0.5cm 0.5cm 0.25cm 0.5cm, clip, width=\columnwidth]{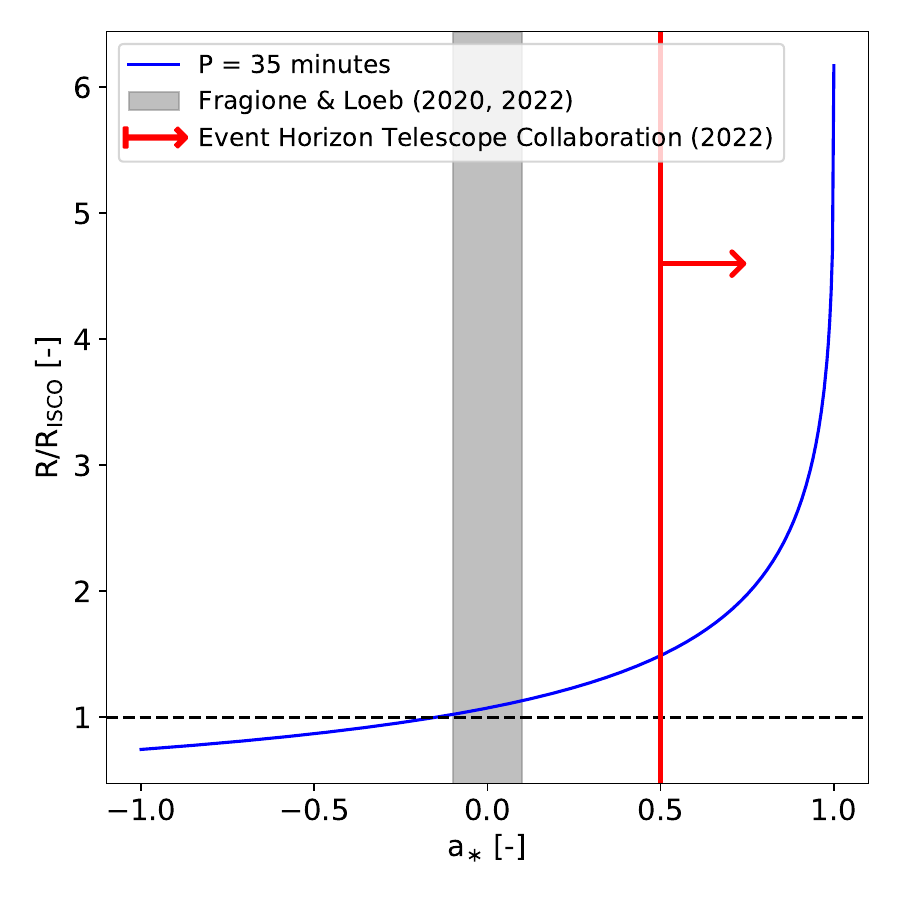}
                \caption{Radius (relative to the innermost circular stable orbit; ISCO) of a 35-minute orbital period as a function of the normalized spin parameter ($a_{\ast}$) for Sgr A* (blue). The gray shaded region denotes the range of normalized spins from dynamical considerations \citep{Fragione2020,Fragione2022}. The red line shows the rough lower limit of Sgr A*'s spin comparing GRMHD simulations with event-horizon scale imaging \citep{EHT2022a,EHT2022e}.}
                \label{fig:isco_plot}
            \end{figure}
            
            The transient $\sim35$-minute peak (Figure \ref{fig:alldata_lomb_scargles}) corresponds to similar timescale detections in near-IR \citep{Gravity2018} and submm \citep{Iwata2020} analyses of Sgr A*'s light curves. \citet{Gravity2018} observed a changing polarization angle consistent with an orbiting hotspot having period $P=40\pm8$ minutes. Similarly, \citet{Iwata2020} used several epochs of ALMA light curves ($230$ GHz) and found $\sim35$-minute quasi-periodic variations at $>5\sigma$ statistical significance. In the context of a non-spinning Schwarzschild black hole ($a_{\ast} = 0$), these results are consistent with variability occurring near the ISCO of Sgr A*. However, there is tension between constraints in Sgr A*'s spin. \citet{Fragione2020, Fragione2022} argued for a low spin ($\left|a_{\ast}\right| \lesssim 0.1$) using the dynamics of the S-star population orbiting Sgr A*. The S-stars reside in two orthogonal disks centered on Sgr A* \citep{Ali2020}; on timescales longer than $\sim1$ Myr, the Lense-Thirring Effect should have randomized these orbits if $|a_{\ast}|\approx 0.5$. This is in stark contrast to \citet{EHT2022a, EHT2022e}, who preferred $a_{\ast} \gtrsim 0.5$ when comparing event-horizon scale images to general relativistic magnetohydrodynamic (GRMHD) simulations of Sgr A*.

            Due to this discrepancy, we cannot definitively determine the orbital radius where a $\sim35$ minute periodicity would originate. Instead, we calculate the orbital radius corresponding to a $35$-minute period for the full range of $a_{\ast}$ and show it in Figure \ref{fig:isco_plot}. We use the period-radius relationship given in \citet[Appendix B]{Gravity2018}:
                \begin{equation}
                    P~\left[\rm{min}\right] = 2.137~\left(\dfrac{M}{4.14\times10^6~M_\odot}\right)~\left(a_{\ast} + \left(\dfrac{2R}{R_{\rm{S}}}\right)^{3/2}\right).
                \end{equation}
            The ISCO radius is given by
                \begin{subequations}
                    \begin{align}
                        R_{\rm{ISCO}} &= \dfrac{R_{\rm{S}}}{2}\left(3 + z_2 \mp \sqrt{\left(3-z_1\right)\left(3 + z_1 + 2z_2\right)}\right);\\
                        z_1 &=  1 + \sqrt[3]{1 - a_{\ast}^2} \left(\sqrt[3]{1  + a_{\ast}} + \sqrt[3]{1 - a_{\ast}}\right),\\
                        z_2 &= \sqrt{3a_{\ast}^2 + z_1^2},
                    \end{align}
                \end{subequations}
            where $\mp$ correspond to prograde and retrograde orbits, respectively \citep{Bardeen1972}. If this quasi-periodicity is caused by orbital motion, the hotspot is on a stable orbit ($R \geq R_{\rm{ISCO}}$) given $a_{\ast}\gtrsim-0.15$. Viscous timescales are unlikely to be the cause as the accretion flow would require a scale height (H/R) $\sim2.2$ for a 35-minute viscous timescale at $3~R_{\rm{S}}$ \citep[see][]{Dexter2014}.

            Several non-orbital causes may be the origin of this structure. \citet{Boyce2022} noted this periodic structure in their analysis of the \Sp\ data but argued it may be formed in processes that can be modeled by correlated red noise. \citet{Witzel2012} and \citet{Subroweit2017} found Sgr A*'s submm variability is also consistent with red noise. If this periodic feature was a statistical product of Sgr A*'s red noise fluctuations, observing it in the submm and IR strongly indicates the variability at these two frequency regimes originates from a single electron population.
            
            However, physical causes are shown in GRMHD modeling. \citet{Ripperda2020} found magnetic reconnection within magnetically-arrested disks (MAD), such as those around Sgr A*, can form ``plasmoid'' hotspots at periodic intervals, and the period of their formation depends on the reconnection rate. Such plasmoids undergo physical expansion, causing a decrease in their average internal magnetic field strength \citep[e.g.,][]{Nathanail2020}. Additionally, \citet{Porth2021} and \citet{Chatterjee2022} studied orbiting flux tubes, regions within the accretion flow with a decreased magnetic field strength compared to the local environment, which could also produce periodic signatures in light curves.
            
            Plasmoids are similar to adiabatically-expanding hotspots, which describe time-delayed emission at different frequencies as the change in optical depth as the hotspot expands. In Figure \ref{fig:alma_spitzer_gp}, we detect a $\sim5$ minute time delay detected between the \Sp\ and ALMA light curves. Taken together, this feature may have been formed by a series of magnetic reconnections within the accretion flow, whose subsequent dynamic expansion caused time-delayed emission between these two frequency regimes. However, plasmoids cause dips in the submm flux from depleted magnetization in the accretion flow rather than a flare occulting it (see Section \ref{ssec:occultation} below). If such events are recorded at different wavelengths, the predicted multiwavelength signatures may be able to distinguish between the two models.
    
        \subsubsection{Occultation Modeling}\label{ssec:occultation}
            \begin{figure*}
                \centering
                \includegraphics[width=\textwidth]{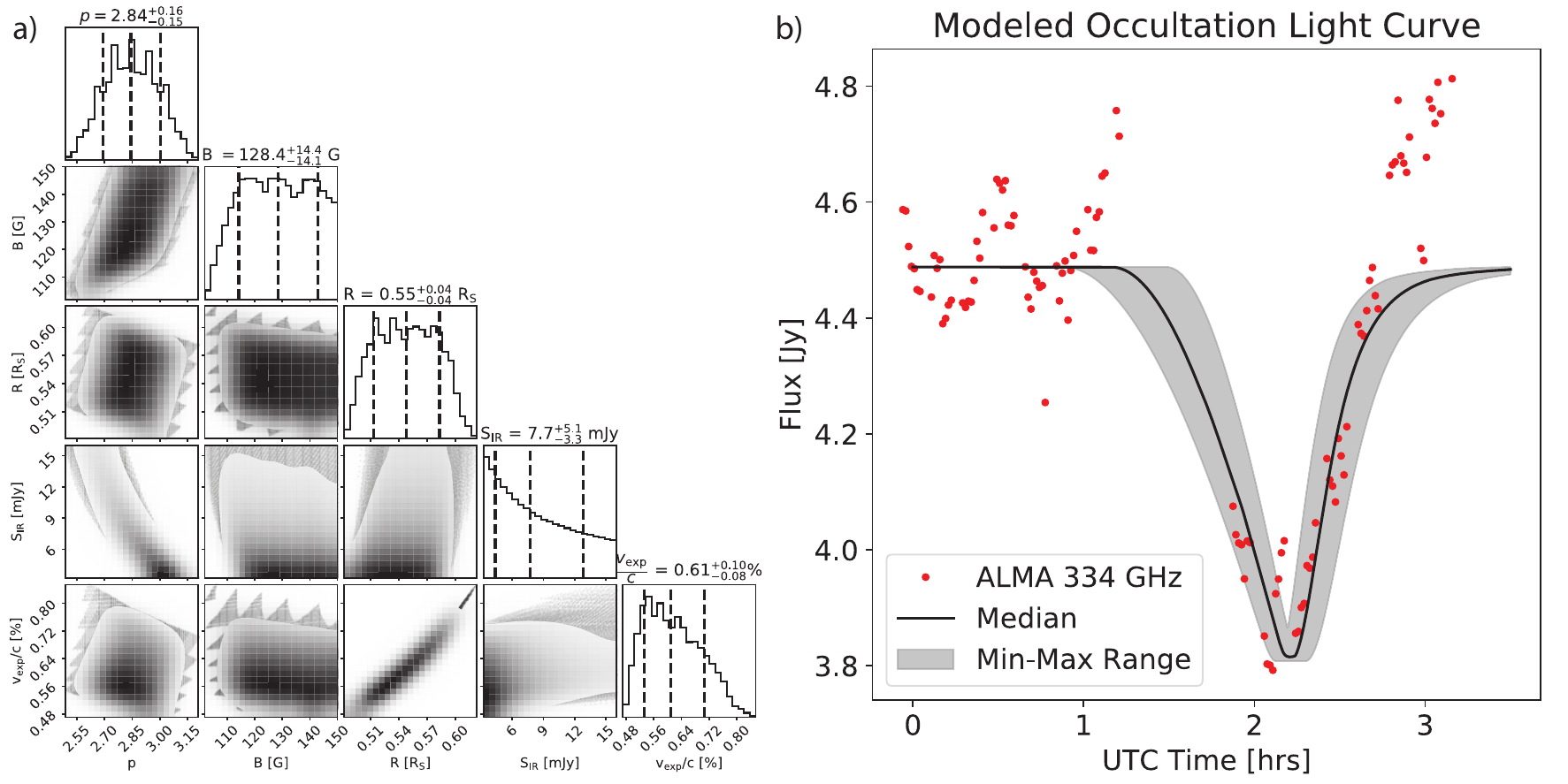}
                \caption{(a): Corner plot showing the physical and observable parameters of the constrained models. The diagonal plots are histograms of the four parameters. We denote the 16th, 50th, and 84th quantiles and are labeled at the top of each plot. The off-diagonal plots are the two-dimensional histograms between the pairs of parameters. (b): The range of models consistent with the occultation. The black line shows the median model flux during the event, while the shaded region shows the total modeled flux range.}
                \label{fig:occultation_cp}
            \end{figure*}
            
            Assuming that there was an IR flare (Figure \ref{fig:july26_xcorr}), Equations \ref{eq:J_nu}--\ref{eq_nE} could be manipulated to solve for the equipartition magnetic field strength, allowing us to directly fit the submm light curve with the occultation model. However, the \Sp\ data during the period of interest is low SNR, and no discernible flare was detected. While an X-ray flare was detected, the exact mechanism producing this emission is still an open question \citep[see, for example,][]{Markoff2001, FYZ2006a, DoddsEden2009, Ponti2017, Abuter2021}. Therefore, we cannot accurately extrapolate the IR flaring flux. Additionally, there is an insufficient range in bandwidth for us to use all four ALMA light curves to directly fit the data. 
            
            Instead, we create a grid of initial physical conditions to determine the IR and submm properties of the plasma. We make grid of $500^3$ realizations linearly sampling the electron energy index ($p\in\left[0.34, 5\right]$), the magnetic field strength ($B\in\left[0.1, 150\right]$ Gauss), and the initial radius ($R\in\left[0.01, 10\right]~\rm{R}_{\rm{S}}$). For each model, we calculate the optically thin IR flux ($S_{\text{IR}}=\Omega J_{\nu} \tau_{\nu}$) and the (absolute) maximum change in the flux at 334 GHz flux (Equation \ref{eq:total}). For this calculation, we assume $S_q = 4.48$ Jy, which is the average flux before the occultation (Figure \ref{fig:july21_xcorr}d). Determining the maximum change in the submm flux is analytically complicated as most of the parameters vary with the radius of the plasma. Therefore, we numerically simulate the light curve of each realization by varying the plasma size between $1-10\times$ its initial value and determine the radius where the (absolute) maximum submm flux change occurs.

            We limit the possible models using several constraints. First, we require the size of the occulting plasma at the maximum change to be larger than its initial size; this constraint stems from the non-zero time delay between the X-ray flare and the submm dip (Figure \ref{fig:july21_xcorr}). Next, we require the change in the submm flux to be negative. Note the submm flux may increase \textit{or} decrease depending on the model conditions. An increase in the submm flux occurs when the plasma's brightness temperature is greater than the accretion flow's, producing a normal submm flare (Section \ref{ssec:flaring_model}). From Figure \ref{fig:alma_spitzer_gp}, we calculate a flux drop of $\sim610$ mJy at 334 GHz but consider models where the flux difference is within $20\%$ of this value ($488-732$ mJy) to account for scatter within the ALMA light curve. 
            
            While we do not detect the IR flare, we can place statistical upper- and lower limits on its flux. As an upper limit, we calculate $5\times$RMSE (RMSE $\sim3.2$ mJy) of the \Sp\ data during the low SNR period (about 1-2.1 hr UTC). The corresponding IR flare hidden within the noise might increase the RMSE, affecting our upper limit. However, since we do not attempt a true fit to the data, we do not revise this upper limit. The lower limit comes from \citet{Witzel2021}, who completed a joint-variability analysis of Sgr A* from submm to X-ray frequencies. They determine each \Sp\ flux measurement has $\sim0.65$ mJy measurement noise. Since \Sp\ was unlikely to measure flares dimmer than this, we use $5\times$ this value as our lower limit on the IR flux. 
            
            We find only $\sim0.02\%~(27533)$ of our initial conditions are consistent with the data. A corner plot of the initial conditions is shown in Figure \ref{fig:occultation_cp}. While we cannot directly fit the light curves, we use the realizations consistent with the data to determine the statistical properties of the flare. We find the drop is consistent with a plasma of initial radius $0.55$ R$_{\rm{S}}$ threaded by a uniform $\sim130$ Gauss magnetic field expanding at $\sim0.6\%$ of the speed of light. This emission is produced by an electron spectrum with index $p\sim2.8$, and we estimate it yields a peak $4.5~\mu$m flare flux $\sim8$ mJy.

            The ``pseudo-posterior'' magnetic field distribution (histograms along the diagonal) is roughly flat, suggesting that we have not probed the best field strengths consistent with the data. This motivates us to test additional models, where we set an unrealistic maximum strength of $500$ Gauss. Even at these strengths, the distribution of values consistent with the data is still flat (not shown). The average magnetic field strength probed by this initial analysis is stronger than typical values found in observation \citep[e.g., $\sim10-100$ G;][]{Ponti2017, Abuter2021, Michail2021c, Michail2021b, Witzel2021} and MHD simulations \citep[e.g., $\sim100$ G;][]{Ressler2020a}. However, we note \citet{Ressler2020b} have shown that average magnetic field strengths above $200$ Gauss are possible near the event horizon (according to full-GRMHD modeling).

            In the context of the given observation, a strong magnetic field and flat distribution is not surprising. Occultations occur when the brightness temperature of the responsible hotspot is lower than that of the accretion flow. The brightness temperature for synchrotron emission scales as $T_B\propto J_\nu / \left(\Omega\nu^2\right)$, where $J_\nu$ is the synchrotron source function. $J_\nu$ itself scales as $J_\nu\propto \nu^{2.5}/\sqrt{B}$ (Equation \ref{eq:J_nu}); therefore, $T_B \propto \sqrt{\nu / B}$. This necessarily requires high magnetic field strengths to balance the value of $\nu$ at submm frequencies. Additionally, the flat distribution of magnetic field strengths is an artifact of degeneracy since we only used one frequency to model the occultation. For one given frequency, there are an infinite number of magnetic field strengths that will produce a hotspot with a brightness temperature lower than the surrounding accretion flow. Observing the event at a different band constrains the required magnetic field strength to produce this feature at multiple frequencies, where the accretion flow's brightness temperature is also different. 
            
            Additionally, the elevated magnetic field strength may be caused by our choice of the electron energy spectrum's lower energy bound, which we assume is $1$ MeV. We find a total relativistic electron density of $\sim1.9\times10^{8}$ cm$^{-3}$ between 1-1000 MeV for our parameters. This $\sim40\times$ higher than the electron density found by \citet{Abuter2021} during the flare on 18 July 2019 but similar to the \citet{Boyce2022} value when considering the same flare undergoing adiabatic expansion. We assume magnetic equipartition with the electron energy density while modeling the flare (see Equation \ref{eq_nE} and Appendix \ref{appendix:a}). Electrons responsible for the majority of synchrotron emission at frequency $\nu$ and magnetic field strength $B$ have Lorentz factors near:
                \begin{equation}
                    \gamma = \sqrt{~3578\left(\dfrac{\nu}{\text{GHz}}\right)\left(\dfrac{\text{Gauss}}{B}\right)}
                    \label{eq:gamma_likely}
                \end{equation}
            \citep{CondonRansom2016}. For 334 GHz and 130 Gauss, $\gamma\sim100$ corresponds to an electron energy $\sim50$ MeV. If we grossly underestimate the lower energy bound of the electron spectrum, this overestimates the electron energy density and the magnetic field strength. By way of example, changing the lower energy bound to $\gamma=10$ ($5$ MeV) for an electron spectrum with $p=2.84$ decreases the electron energy density by a factor $\sim4$, reducing the magnetic field strength by a factor $\sim2$ ($B \propto \sqrt{U_e}$). 
            
            Obtaining at least one light curve in the radio would constrain the lower energy bound of the electron spectrum ($E_1$). The dip in the radio flux is lessened \citep[since the intrinsic accretion flow size increases at lower frequencies, decreasing $f$;][]{Cho2022}. However, the time delay of the dip relative to submm depends on the critical optical depth of the plasma, which is solely a function of $p$ (Equation \ref{eq:tau_nu}). If the predicted time delay at radio frequencies did not match the value extrapolated from the submm, it could suggest curvature in the electron spectrum between energies probed at the two frequencies, changing the intrinsic optical depth.

            \subsubsection{Signatures of Occulting Hotspots Offset from Accretion Flow}
                Our definition of $f$ implies the occulting hotspot is fully embedded in the accretion flow at any frequency, which is a central assumption in our model and affects the physical parameters. In Figure \ref{fig:occultation_diagram}, we show both possibilities where the flare is fully and partially embedded in the accretion flow. In the concentric case, the total observed flux of the system is $S_q(\nu)$ (number 1) before the flare forms. As the plasma expands, it is optically thick and occults part of the accretion flow (number 2) until it reaches the critical optical depth (number 3). The continued expansion causes the emission to become optically thin, lessening the extinction the accretion flow undergoes (number 4) until only the flow remains (number 5).
    
                In the offset model (bottom panels), an interesting feature occurs. As in the previous model, the total emission measured is the accretion flow flux (number 1). However, as the plasma begins to expand, it does not occult any part of the flow, raising the overall flux of the system (number 2). Eventually, the hotspot's size is large enough, and its optical depth is sufficient to begin shadowing the accretion flow (number 3). As in the previous case, the continuing expansion will cause a dip in the light curve (number 4) and eventually will fade away until only the accretion flow remains (number 5). We note such offset flares require a precise combination of distance from the accretion flow and optical depth for the occultation feature to occur. Additionally, if the hotspot's distance from the accretion flow is sufficiently far, the hotspot becomes optically thin before it causes an occultation, rendering only a flare in the light curve. Therefore, the offset also explains the lack of occultation events during every flaring event.  
    
                The frequency-dependent size of Sgr A* may also play a role in determining which occultation profile occurs. At radio frequencies, the intrinsic size of this source is larger than the reported hotspot sizes. \citet{Cho2022} found Sgr A*'s intrinsic extent to be $\sim30~R_{\rm{S}}\times23~R_{\rm{S}}$ at 43 GHz. Using the relationship given in their paper, the 334 GHz size of Sgr A* is $\sim2.3~R_{\rm{S}}\times1.7~R_{\rm{S}}$, comparable to the hotspot sizes. The Gaussian-process-interpolated ALMA light curve (Figure \ref{fig:alma_spitzer_gp}) suggests this offset profile was detected on 21 July 2019. \citet{Marrone2008} also may have detected a similar case on 17 July 2006 during simultaneous X-ray and submm observations (see their Figure 3). They saw increases in the $850~\mu$m and $1.3$ mm light curves at the beginning of an X-ray flare, then reached a minimum flux near its peak. Future multiwavelength campaigns should observe complementary radio frequencies to test the frequency-dependent nature of this occultation model.
    \\\\
    \subsection{26 July 2019}\label{ssec:flare_modeling}
        
        \begin{figure}
            \centering
            \includegraphics[width=\columnwidth]{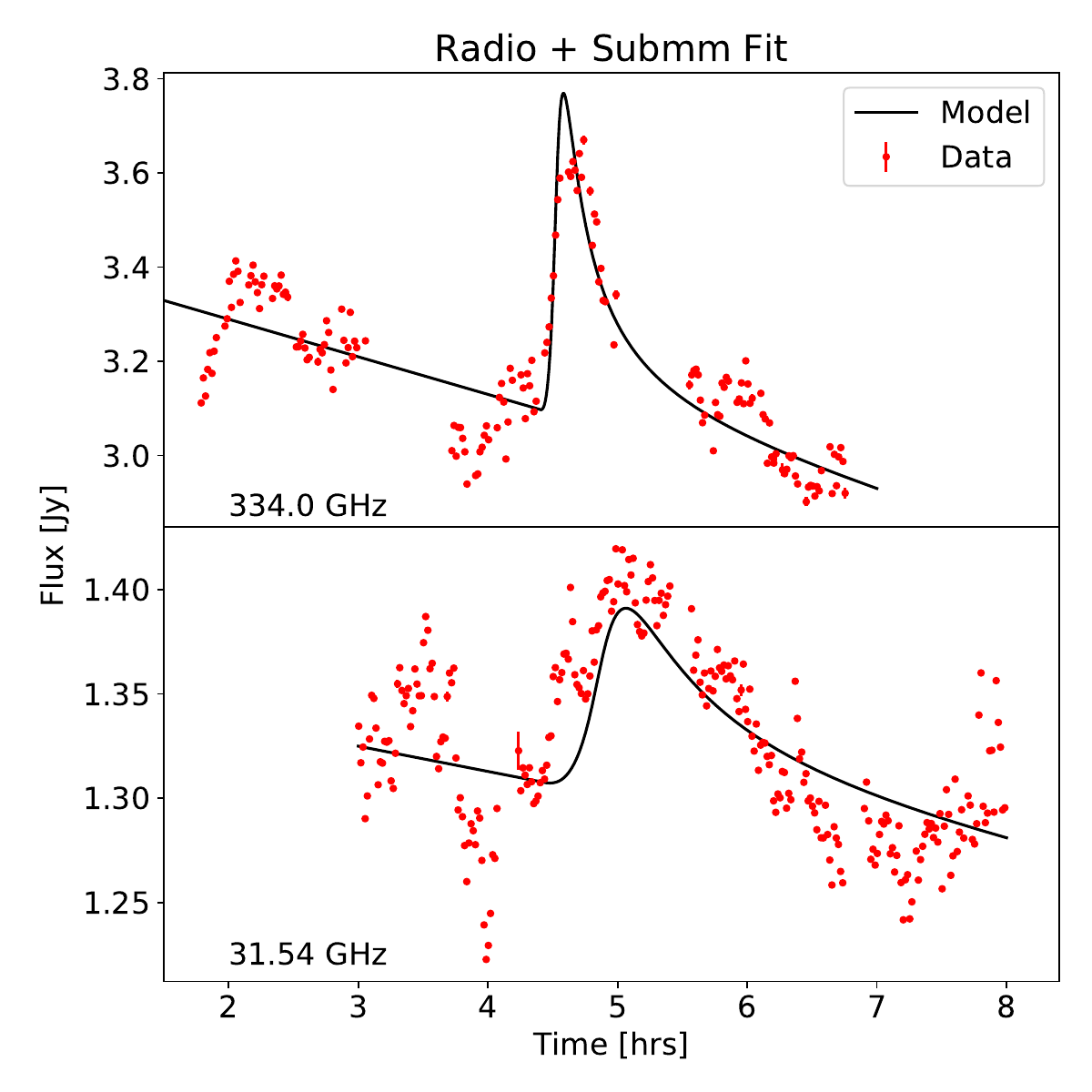}
            \caption{The joint radio-submm adiabatic-expansion hotspot fits. The red points are the observed emission, while the black lines are the modeled light curves during the fitted ranges. The corresponding fit parameter values are given in Table \ref{tab:plasmon_fits}.}
            \label{fig:plasmon_fits}
        \end{figure}

        Using the adiabatic-cooling model described in Section \ref{ssec:flaring_model}, we jointly fit the radio and submm light curves shown in Figure \ref{fig:july26_xcorr}. Modeling the IR light curve is complicated as synchrotron cooling dominates adiabatic cooling at these high frequencies. For example, the synchrotron cooling timescale at $4.5~\mu$m in a 50 G field is $\sim6$ minutes, while the adiabatic cooling is at $\sim$hourly timescales. Attempting to fit all three frequency light curves simultaneously requires a more substantial model to account for both adiabatic and synchrotron cooling, which is beyond the scope of this analysis. Therefore, we do not model the IR light curve, as the modeled parameters will have little physical meaning. Instead, they will be over-biased such that the ``best-fit'' adiabatic expansion picture mimics a synchrotron cooling profile.

        The VLA and ALMA light curves have sufficient spectral windows to simultaneously fit the flare and quiescent emission. The flaring emission is fit using a single set of hotspot parameters. On the other hand, we use two different sets of quiescent parameters to independently model the radio and submm emission, which is necessary for two reasons. First, the quiescent model includes a phenomenological parameter to model the time-slope of the accretion flow at each frequency range. As each band probes different radii within the accretion flow \citep{Cho2022}, there is no guarantee these values will be equivalent. Second, Sgr A*'s spectral index flattens from radio to submm frequencies \citep[i.e.,][]{Bower2015}, which would require fitting three spectral indices (intra-radio, intra-submm, and radio-submm). By fitting one quiescent model per frequency band, we can determine the intra-radio and intra-submm indices and then infer the one from radio to submm using the quiescent flux densities at each reference frequency (i.e., $I_0^\nu$, Equation \ref{eq:quiescent}). The result of this joint fit is shown in Figure \ref{fig:plasmon_fits}, and the parameter values are given in Table \ref{tab:plasmon_fits}.
    
        \begin{deluxetable*}{cccc}
            \tabletypesize{\normalsize}
            \tablewidth{\textwidth}
            \tablenum{1}
            \tablecolumns{4}
            \tablecaption{Fitted parameters for the joint submm and radio light curve modeling shown in Figure \ref{fig:plasmon_fits}. Quoted errors are given at $1\sigma$ significance.\label{tab:plasmon_fits}}
            \tablehead{Parameter & Description & Unit & Value}
            \startdata
               \multicolumn{4}{c}{Hotspot Parameters}\\\hline
               $S_0$ & Peak Flare Flux at 334 GHz & mJy & $686 \pm 6$\\
               $t_0$ & Flare UTC Peak Time at 334 GHz & hr UTC & $4.58 \pm 0.01$\\
               $p$ & Electron Spectrum Index & -- & $0.59 \pm 0.01$ \\
               $v_\text{exp}$ & Normalized Radial Expansion Speed & hr$^{-1}$ & $5.36 \pm 0.05$ \\\hline
               \multicolumn{4}{c}{Radio Quiescent Parameters}\\\hline
               $I_0^{\rm{radio}}$ & Quiescent Flux at 37.54 GHz at $t=t_0$ & mJy & $1305.6\pm0.6$ \\
               $\delta I_t^{\rm{radio}}$ & Quiescent Time-Slope at 37.54 GHz & mJy hr$^{-1}$ & $-12.2\pm0.3$\\
               $\alpha_{\rm{radio}}$ & Ka-band Quiescent Spectral Index & -- & $0.074 \pm 0.003$ \\\hline
               \multicolumn{4}{c}{Submm Quiescent Parameters}\\\hline
               $I_0^{\rm{submm}}$ & Quiescent Flux at 334 GHz at $t=t_0$ & mJy & $3083\pm2$ \\
               $\delta I_t^{\rm{submm}}$ & Quiescent Time-Slope at 334 GHz & mJy hr$^{-1}$ & $-79.9 \pm 0.1$ \\
               $\alpha_{\rm{submm}}$ & Band 7 Quiescent Spectral Index & -- & $-0.099 \pm 0.023$ \\\hline
            \enddata
        \end{deluxetable*}

        \subsubsection{Quiescent Properties}
            We find a spectral index that is roughly flat ($\alpha\sim0.07$) at Ka-band. \citet{Herrnstein2004} find a variable spectral index at similar wavelengths that range from $\sim-0.1$ to $0.7$. We note, however, that their observations do not account for flaring emission and are an ensemble average of both the quiescent and variable components. The (ensemble-average) spectral energy distribution of Sgr A* from \citet{Brinkerink2015} suggested a similarly flat spectrum and comparable flux density at radio frequencies as ours here. Similarly, our fitted submm spectral index and flux are well within the range of those observed \citep[e.g.,][]{Wielgus2022a}. Using the fitted quiescent fluxes, we also determine the average spectral index between 31.54 and 334 GHz is $\alpha\sim0.36$, which is consistent with the canonical value of $\alpha=0.33$ for optically-thick accretion flows \citep{Pringle1981} and those in the literature \citep[e.g.,][]{Michail2021b}.
    
            The consistency between our fitted quiescent emission in 2019 and previous analyses is somewhat shocking. IR variability suggests that Sgr A* was more variable in 2019 than in the past $\sim20$ years \citep{Do2019, Weldon2023}, which they concluded occurs from enhanced accretion onto the event horizon. As flaring events occur within the accretion flow, one would expect similar outliers in the quiescent submm and radio emission. However, this analysis does not show similar drastic changes, even at 334 GHz, where the emission primarily occurs near the event horizon, agreeing with \citet{Murchikova2021}, who find no change in either the 230 GHz flux or variability statistics in the aftermath of this bright IR flare. Additional radio observations of Sgr A* during 2019 should be analyzed to determine if a similar change in the radio continuum properties occurred.

        \subsubsection{Hotspot Properties}\label{ssec:hotspot_properties}
            The first four rows in Table \ref{tab:plasmon_fits} list the fitted parameters for the hotspot. The plasma produces a $686$ mJy flare at 334 GHz from a shallow electron energy spectrum with an index $p=0.59$. The low value of $p$ is caused by neglecting synchrotron cooling in the adiabatically-expanding hotspot model. Values typically range from $p\approx1-3$ \citep[i.e.,][]{FYZ2009, Eckart2009, Michail2021c, Michail2023, Witzel2021, Boyce2022}, where $p=1$ indicates an extremely strong and efficient shock \citep{Bell1978a, Bell1978b}. We discuss the low value of $p$ in Section \ref{ssec:synch_cool}. 
            
            Given these values, we can calculate an equipartition magnetic field strength and radius for the observed flare. Once again, we assume the $E^{-0.59}$ electron energy spectrum is valid from energies between 1 to 1000 MeV consistent with Section \ref{ssec:occultation}. We determine the hotspot to have a radius $0.47$ $R_{\text{S}}$ at the time of the peak submm emission, which is threaded by a magnetic field $\sim187$ Gauss. The mass is $3.7\times10^{18}$ grams, and the electron density is $2.82\times10^6$ cm$^{-3}$. Assuming linear expansion of the hotspot yields an expansion velocity of $0.029c$. This combination of magnetic field strength and electron number density is similar to the time-averaged conditions within a few $R_{\text{S}}$ of a strongly-jetted accretion flow around Sgr A*, as modeled with GRMHD \citep{Chatterjee2021}.
            
            \citet{Subroweit2017} complete a statistical analysis of 100 and 345 GHz light curves of Sgr A*. By relating the flaring flux distribution with the adiabatically expanding hotspot model, they suggest the majority of flares have size $\sim0.67~R_{\text{S}}$ and expand with ``intermediate'' velocities of order $0.01c$. The expansion velocities have generally been found to lie in the range of $0.001-0.1c$ \citep[e.g.,][]{FYZ2008}. Similarly, the inferred radius of these hotspots are order $\sim R_{\text{S}}$, with most of the GRAVITY IR interferometry results preferring $0.5-1.5~R_{\text{S}}$ \citep[e.g.,][]{Gravity2020, Gravity2020orbit, Abuter2021}.
            
            As with the models consistent with the occultation event, the magnetic field strength here is much stronger than previously observed by a factor $\sim2$. This might be attributed to the incorrect upper energy bound rather than the lower one described during the occultation. The sensitivity of the magnetic field strength on the upper or lower bound depends on the electron energy index. Equation \ref{eq:ue} gives the electron energy density as a function of $p$, $E_{2}$, and $E_{1}$ (the maximum and minimum electron energies, respectively). For $p>2$ (in the case of the occultation), the value of $U_e$ is more sensitive to $E_1$ instead of $E_2$ since $E_2^{2-p} < E_1^{2-p}$. For $p\in[0, 2)$, the opposite is true as in our case of $p=0.59$, $U_e\propto E_2^{1.41}-E_1^{1.41}$. Using Equation \ref{eq:gamma_likely} for 334 GHz at 180 Gauss yields $\gamma \approx 80$ (an electron energy of $40$ MeV). Reducing the original upper energy bound to 175 MeV \citep[$\gamma = 350$, consistent with the modeling in][]{Boyce2022} yields an equipartition magnetic field strength $94$ Gauss, electron density $3.8\times10^{6}$ cm$^{-3}$, mass $2.9\times10^{18}$ grams, radius $0.39~R_{\text{S}}$, and expansion speed $0.024c$.
            
    \subsection{Effects of Synchrotron Cooling}\label{ssec:synch_cool}
        Throughout the models presented in Section \ref{sec:models}, we have ignored any contribution of synchrotron cooling to the light curves. However, at the submm frequencies probed here, it somewhat affects the observed profiles. In the occultation modeling (Section \ref{ssec:occultation}), the adiabatic and synchrotron cooling times are roughly equal. If we define the adiabatic timescale as $t_{\text{ad}}=R_{\text{max}}/v_{\text{exp}}$, we find this value to be $\sim 1$ hour, roughly double the synchrotron cooling timescale at 334 GHz in a 130 Gauss field, $t_{\text{sc}}\sim0.5$ hr. Ignoring this effect will tend to over-predict the magnetic field strength as the electrons will have lower characteristic energies than calculated via adiabatic expansion alone. Since the characteristic electron energy $E_\nu\propto B^{-1/2}$ (Equation \ref{eq:e_nu}), the fitting results will tend to bias toward higher magnetic field strengths (lower energy) to account for this discrepancy.
        
        Likewise, the effect of synchrotron cooling is visible in the submm light curve in Figure \ref{fig:plasmon_fits}, as the predicted adiabatic model peak flux is higher than that of the observed value. This also tends to increase the equipartition magnetic field strength as above; however, it also affects the fitted value of $p$. At the peak submm emission, a sufficient proportion of electrons have cooled faster via adiabatic \textit{and} synchrotron processes, which decreases the optical depth of the plasma, suppressing the peak flux as prescribed by the adiabatic picture. 
        
        As mentioned in Section \ref{ssec:hotspot_properties}, the low value of $p$ measured in the joint radio-submm plasmon fit is caused by ignoring synchrotron cooling in the evolution of the submm flare. The relationship between the peak flux of an adiabatically-expanding hotspot at two different frequencies is $S_{\nu_2} = S_{\nu_1}\left(\nu_2/\nu_1\right)^{\frac{7p+3}{4p+6}}$ \citep{VDL1966}. As discussed above, synchrotron cooling will tend to decrease the peak flux where it is comparable or dominant to adiabatic cooling. $p$ will be biased towards lower values to fit this dimmer-than-expected submm peak with the radio light curves. To confirm this, we independently fit the radio and submm data with a single adiabatically-cooling hotspot (not shown). We find $p\sim1.3$ for both the radio and submm fits. By way of example, assume a $100$ mJy flare occurs at $31.54$ GHz (as estimated from Figures \ref{fig:july26_xcorr}d and \ref{fig:plasmon_fits}). For $p=0.6$ and $p=1.3$, we find peak $334$ GHz fluxes of $757$ and $1282$ mJy, respectively. The lower value of $p$ better models the observed submm light curve; however, by ignoring synchrotron cooling, we have effectively obtained an unrealistic value of $p$.
        
        Overall, ignoring synchrotron cooling causes a biased value of $p$ and tends to overpredict the calculated equipartition magnetic field strength compared to previous observational and theoretical values. Future analyses using this adiabatic cooling model (especially when synchrotron cooling is dominant to adiabatic cooling) need to account for this additional effect. We note synchrotron cooling is not the only missing process throughout this work. For example, the lack of general relativistic effects, such as lensing, may also affect our results. However, including these additional processes is outside the scope of this current work.
        
\section{Summary and Conclusions}\label{sec:conclusions}

    We present and analyze the data during the final two days of a multiwavelength campaign to observe Sgr A*, which occurred in July 2019 and was the last such campaign with the \Sp~\textit{Space Telescope}. Analysis of the first day was presented in \citetalias{Michail2021c}. The participating instruments ranged in energy from radio to X-ray energies. The observations showed Sgr A* to be remarkably active. On 26 July, we observed a flare in the IR, submm, and radio, which presents the most evident causal link between the variability mechanisms in these frequency regimes. We model the radio and submm light curves using an adiabatically-expanding synchrotron hotspot. We find a transient $\sim35$ minute periodicity in the \Sp\ light curve on 21 July, corresponding to slightly time-delayed peaks observed with ALMA, suggesting a hotspot on a stable orbit around Sgr A*. We detect a drop in the submm flux shortly after an X-ray flare on the same day. We modeled this feature using an adiabatically-expanding synchrotron hotspot, which occults the accretion flow during its dynamic expansion. We present the model in this paper after generalizing it from a specific case for $p=2$ \citep{FYZ2010}. 
    
    During both days, we find the magnetic field strength estimates are uncertain because of the lack of constraints in the upper and lower energy bounds of the electron spectrum and the omission of synchrotron cooling. The absence of lower frequency radio light curves does not allow for a robust lower energy bound. Synchrotron cooling plays a crucial role in the profiles of the submm light curves. The magnetic field strengths we estimate are uncertain on both days because of the lack of data constraining the bound of the electron energy spectrum and neglecting synchrotron cooling, which will tend to increase the equipartition magnetic field strength required to model the observed features. Modeling flares using adiabatically-expanding synchrotron hotspots with future high-frequency submm and IR observations will require accounting for synchrotron cooling losses during the expansion phase, placing a strong bound on the upper energy limit. Supplementing high-quality radio frequency light curves with such observations will constrain the bounds of the electron spectrum and better bound possible values of the magnetic field strength responsible for the emission.
    \\\\

\section*{Acknowledgments}
    We are grateful to Sean Carey, Dan Marrone, and Steven Willner for useful comments that strengthened this manuscript. We additionally thank the referee for their very helpful and constructive feedback. This work is partially supported by NSF grant AST-2305857 and through SOSP21A-003 and SOSPADA-026 from the NRAO. SM is grateful for support from the Dutch Research Council (NWO) VICI award (grant 639.043.513) and the European Research Council (ERC) SyG ``BlackHolistic'' (grant 101071643). GGF, JLH, and HAS acknowledge support for this work from the NASA ADAP program under NASA grant 80NSSC18K0416. The National Radio Astronomy Observatory is a facility of the National Science Foundation operated under cooperative agreement by Associated Universities, Inc. This paper makes use of the following ALMA data: ADS/JAO.ALMA\#2018.A.00050.T. ALMA is a partnership of ESO (representing its member states), NSF (USA) and NINS (Japan), together with NRC (Canada), MOST and ASIAA (Taiwan), and KASI (Republic of Korea), in cooperation with the Republic of Chile. The Joint ALMA Observatory is operated by ESO, AUI/NRAO and NAOJ. Based on observations made with the NASA/ESA Hubble Space Telescope, obtained from the data archive at the Space Telescope Science Institute. STScI is operated by the Association of Universities for Research in Astronomy, Inc. under NASA contract NAS 5-26555. This work is based on archival data obtained with the Spitzer Space Telescope, which was operated by the Jet Propulsion Laboratory, California Institute of Technology under a contract with NASA. The scientific results reported in this article are based in part on data obtained from the Chandra Data Archive. This research has made use of software provided by the Chandra X-ray Center (CXC) in the application packages CIAO. This research has made use of data obtained from the Chandra Source Catalog, provided by the Chandra X-ray Center (CXC) as part of the Chandra Data Archive. This research was supported in part through the computational resources and staff contributions provided for the Quest high-performance computing facility at Northwestern University which is jointly supported by the Office of the Provost, the Office for Research, and Northwestern University Information Technology.

%

\vspace{5mm}
\facilities{VLA, ALMA, HST (WFC3), Spitzer, CXO (ACIS)}


\software{\texttt{AIPS},
          \texttt{astropy} \citep{astropy:2013, astropy:2018},
          \texttt{BBlocks} \citep{BBlocks},
          \texttt{CASA} \citep{McMullin2007},
          \texttt{CIAO}     \citep{Ciao2006},
          \texttt{LMFIT}     \citep{LMFIT},
          \texttt{matplotlib} \citep{matplotlib},
          \texttt{numpy}     \citep{numpy},
          \texttt{pandas}      \citep{pandas},
          \texttt{pyCCF} \citep{Peterson1998, PYCCF},
          \texttt{scipy}     \citep{scipy},
          \texttt{sklearn} \citep{sklearn}
          }



\appendix

\section{Deriving the Equipartition and Electron Spectrum Factors}\label{appendix:a}
For an electron distribution $n(E)\propto E^{-p}$ between energies $E_1$ and $E_2$, the electron energy density is:

    \begin{align}
        U_e = \int_{E_1}^{E_2} E~n(E)~dE = \int_{E_1}^{E_2} n_0E\left(\dfrac{E}{E_0}\right)^{-p}~dE.
    \end{align}
    This integration yields
    \begin{align}
        U_e(p) = \begin{cases}
        \dfrac{n_0E_0^2}{2-p}\left(\left(\dfrac{E_2}{E_0}\right)^{2-p} - \left(\dfrac{E_1}{E_0}\right)^{2-p}\right), & p \geq 0, p\neq2, \\\\
        n_0E_0^2\ln\left(\dfrac{E_2}{E_1}\right), & p=2,
        \label{eq:ue}
        \end{cases}
    \end{align}
    where $n_0$ has units of $\rm{erg}^{-1}\rm{ cm}^{-3}$. Writing $U_B$ and $U_e$ as $U_B = \left(U_B/U_e\right)U_e$, plugging in for $U_e$ from Equation \ref{eq:ue}, and substituting $U_B = B^2/8\pi$:
    \begin{align}
        \dfrac{B^2}{8\pi} = 
        \begin{cases}
        \left(\dfrac{U_B}{U_e}\right) \dfrac{n_0E_0^2}{2-p}\left(\left(\dfrac{E_2}{E_0}\right)^{2-p} - \left(\dfrac{E_1}{E_0}\right)^{2-p}\right), & p \geq 0, p\neq2 \\\\
        \left(\dfrac{U_B}{U_e}\right) n_0E_0^2\ln\left(\dfrac{E_2}{E_1}\right), & p=2.
        \end{cases}
    \end{align}
    
    \noindent Solving for $n_0$:
    \begin{align}
        n_0 = \begin{cases}
        \dfrac{B^2}{8\pi E_0^2}\dfrac{2-p}{\left(\dfrac{U_B}{U_e}\right)\left(\left(\dfrac{E_2}{E_0}\right)^{2-p} - \left(\dfrac{E_1}{E_0}\right)^{2-p}\right)}, & p \geq 0, p\neq2
        \\\\
        \dfrac{B^2}{8\pi E_0^2}\dfrac{1}{\left(\dfrac{U_B}{U_e}\right)\ln\left(\dfrac{E_2}{E_1}\right)}, & p=2.
        \end{cases}
    \end{align}
    
    \noindent By defining,
    \begin{align}
        \varphi(p) \equiv \begin{cases}
        \left(\dfrac{U_B}{U_e}\right)\dfrac{\left(\dfrac{E_2}{E_0}\right)^{2-p} - \left(\dfrac{E_1}{E_0}\right)^{2-p}}{2-p}, & p \geq 0, p\neq2 \\\\
        \left(\dfrac{U_B}{U_e}\right)\ln\left(\dfrac{E_2}{E_1}\right), & p=2,
        \end{cases}
    \end{align}
    we obtain
    \begin{equation}
        n(E) = \dfrac{B^2}{8\pi\varphi E_0^2}\left(\dfrac{E}{E_0}\right)^{-p},
    \end{equation}
    which is Equation \ref{eq_nE} above. Due to adiabatic cooling of the ultra-relativistic electron gas, $E_1$ and $E_2$ scale inversely with radius. Therefore, $\varphi(p)\propto R^{2-p}$, which must be properly accounted for while modeling the overall occultation light curves.

\bibliography{campaign2019.bib}{}

\begin{thebibliography}{}
\expandafter\ifx\csname natexlab\endcsname\relax\def\natexlab#1{#1}\fi
\providecommand{\url}[1]{\href{#1}{#1}}
\providecommand{\dodoi}[1]{doi:~\href{http://doi.org/#1}{\nolinkurl{#1}}}
\providecommand{\doeprint}[1]{\href{http://ascl.net/#1}{\nolinkurl{http://ascl.net/#1}}}
\providecommand{\doarXiv}[1]{\href{https://arxiv.org/abs/#1}{\nolinkurl{https://arxiv.org/abs/#1}}}

\bibitem[{{Ali} {et~al.}(2020){Ali}, {Paul}, {Eckart}, {Parsa}, {Zajacek}, {Pei{\ss}ker}, {Subroweit}, {Valencia-S.}, {Thomkins}, \& {Witzel}}]{Ali2020}
{Ali}, B., {Paul}, D., {Eckart}, A., {et~al.} 2020, \apj, 896, 100, \dodoi{10.3847/1538-4357/ab93ae}

\bibitem[{{Astropy Collaboration} {et~al.}(2013){Astropy Collaboration}, {Robitaille}, {Tollerud}, {Greenfield}, {Droettboom}, {Bray}, {Aldcroft}, {Davis}, {Ginsburg}, {Price-Whelan}, {Kerzendorf}, {Conley}, {Crighton}, {Barbary}, {Muna}, {Ferguson}, {Grollier}, {Parikh}, {Nair}, {Unther}, {Deil}, {Woillez}, {Conseil}, {Kramer}, {Turner}, {Singer}, {Fox}, {Weaver}, {Zabalza}, {Edwards}, {Azalee Bostroem}, {Burke}, {Casey}, {Crawford}, {Dencheva}, {Ely}, {Jenness}, {Labrie}, {Lim}, {Pierfederici}, {Pontzen}, {Ptak}, {Refsdal}, {Servillat}, \& {Streicher}}]{astropy:2013}
{Astropy Collaboration}, {Robitaille}, T.~P., {Tollerud}, E.~J., {et~al.} 2013, \aap, 558, A33, \dodoi{10.1051/0004-6361/201322068}

\bibitem[{{Astropy Collaboration} {et~al.}(2018){Astropy Collaboration}, {Price-Whelan}, {Sip{\H{o}}cz}, {G{\"u}nther}, {Lim}, {Crawford}, {Conseil}, {Shupe}, {Craig}, {Dencheva}, {Ginsburg}, {Vand erPlas}, {Bradley}, {P{\'e}rez-Su{\'a}rez}, {de Val-Borro}, {Aldcroft}, {Cruz}, {Robitaille}, {Tollerud}, {Ardelean}, {Babej}, {Bach}, {Bachetti}, {Bakanov}, {Bamford}, {Barentsen}, {Barmby}, {Baumbach}, {Berry}, {Biscani}, {Boquien}, {Bostroem}, {Bouma}, {Brammer}, {Bray}, {Breytenbach}, {Buddelmeijer}, {Burke}, {Calderone}, {Cano Rodr{\'\i}guez}, {Cara}, {Cardoso}, {Cheedella}, {Copin}, {Corrales}, {Crichton}, {D'Avella}, {Deil}, {Depagne}, {Dietrich}, {Donath}, {Droettboom}, {Earl}, {Erben}, {Fabbro}, {Ferreira}, {Finethy}, {Fox}, {Garrison}, {Gibbons}, {Goldstein}, {Gommers}, {Greco}, {Greenfield}, {Groener}, {Grollier}, {Hagen}, {Hirst}, {Homeier}, {Horton}, {Hosseinzadeh}, {Hu}, {Hunkeler}, {Ivezi{\'c}}, {Jain}, {Jenness}, {Kanarek}, {Kendrew}, {Kern}, {Kerzendorf}, {Khvalko}, {King}, {Kirkby}, {Kulkarni},
  {Kumar}, {Lee}, {Lenz}, {Littlefair}, {Ma}, {Macleod}, {Mastropietro}, {McCully}, {Montagnac}, {Morris}, {Mueller}, {Mumford}, {Muna}, {Murphy}, {Nelson}, {Nguyen}, {Ninan}, {N{\"o}the}, {Ogaz}, {Oh}, {Parejko}, {Parley}, {Pascual}, {Patil}, {Patil}, {Plunkett}, {Prochaska}, {Rastogi}, {Reddy Janga}, {Sabater}, {Sakurikar}, {Seifert}, {Sherbert}, {Sherwood-Taylor}, {Shih}, {Sick}, {Silbiger}, {Singanamalla}, {Singer}, {Sladen}, {Sooley}, {Sornarajah}, {Streicher}, {Teuben}, {Thomas}, {Tremblay}, {Turner}, {Terr{\'o}n}, {van Kerkwijk}, {de la Vega}, {Watkins}, {Weaver}, {Whitmore}, {Woillez}, {Zabalza}, \& {Astropy Contributors}}]{astropy:2018}
{Astropy Collaboration}, {Price-Whelan}, A.~M., {Sip{\H{o}}cz}, B.~M., {et~al.} 2018, \aj, 156, 123, \dodoi{10.3847/1538-3881/aabc4f}

\bibitem[{{Baganoff} {et~al.}(2001){Baganoff}, {Bautz}, {Brandt}, {Chartas}, {Feigelson}, {Garmire}, {Maeda}, {Morris}, {Ricker}, {Townsley}, \& {Walter}}]{Baganoff2001}
{Baganoff}, F.~K., {Bautz}, M.~W., {Brandt}, W.~N., {et~al.} 2001, \nat, 413, 45, \dodoi{10.1038/35092510}

\bibitem[{{Bardeen} {et~al.}(1972){Bardeen}, {Press}, \& {Teukolsky}}]{Bardeen1972}
{Bardeen}, J.~M., {Press}, W.~H., \& {Teukolsky}, S.~A. 1972, \apj, 178, 347, \dodoi{10.1086/151796}

\bibitem[{{Bell}(1978{\natexlab{a}})}]{Bell1978a}
{Bell}, A.~R. 1978{\natexlab{a}}, \mnras, 182, 147, \dodoi{10.1093/mnras/182.2.147}

\bibitem[{{Bell}(1978{\natexlab{b}})}]{Bell1978b}
---. 1978{\natexlab{b}}, \mnras, 182, 443, \dodoi{10.1093/mnras/182.3.443}

\bibitem[{{Boehle} {et~al.}(2016){Boehle}, {Ghez}, {Sch{\"o}del}, {Meyer}, {Yelda}, {Albers}, {Martinez}, {Becklin}, {Do}, {Lu}, {Matthews}, {Morris}, {Sitarski}, \& {Witzel}}]{Boehle2016}
{Boehle}, A., {Ghez}, A.~M., {Sch{\"o}del}, R., {et~al.} 2016, \apj, 830, 17, \dodoi{10.3847/0004-637X/830/1/17}

\bibitem[{{Bouffard} {et~al.}(2019){Bouffard}, {Haggard}, {Nowak}, {Neilsen}, {Markoff}, \& {Baganoff}}]{Bouffard2019}
{Bouffard}, {\'E}., {Haggard}, D., {Nowak}, M.~A., {et~al.} 2019, \apj, 884, 148, \dodoi{10.3847/1538-4357/ab4266}

\bibitem[{{Bower} {et~al.}(2015){Bower}, {Markoff}, {Dexter}, {Gurwell}, {Moran}, {Brunthaler}, {Falcke}, {Fragile}, {Maitra}, {Marrone}, {Peck}, {Rushton}, \& {Wright}}]{Bower2015}
{Bower}, G.~C., {Markoff}, S., {Dexter}, J., {et~al.} 2015, \apj, 802, 69, \dodoi{10.1088/0004-637X/802/1/69}

\bibitem[{{Bower} {et~al.}(2019){Bower}, {Dexter}, {Asada}, {Brinkerink}, {Falcke}, {Ho}, {Inoue}, {Markoff}, {Marrone}, {Matsushita}, {Moscibrodzka}, {Nakamura}, {Peck}, \& {Rao}}]{Bower2019}
{Bower}, G.~C., {Dexter}, J., {Asada}, K., {et~al.} 2019, \apjl, 881, L2, \dodoi{10.3847/2041-8213/ab3397}

\bibitem[{{Boyce} {et~al.}(2019){Boyce}, {Haggard}, {Witzel}, {Willner}, {Neilsen}, {Hora}, {Markoff}, {Ponti}, {Baganoff}, {Becklin}, {Fazio}, {Lowrance}, {Morris}, \& {Smith}}]{Boyce2019}
{Boyce}, H., {Haggard}, D., {Witzel}, G., {et~al.} 2019, \apj, 871, 161, \dodoi{10.3847/1538-4357/aaf71f}

\bibitem[{{Boyce} {et~al.}(2022){Boyce}, {Haggard}, {Witzel}, {Fellenberg}, {Willner}, {Becklin}, {Do}, {Eckart}, {Fazio}, {Gurwell}, {Hora}, {Markoff}, {Morris}, {Neilsen}, {Nowak}, {Smith}, \& {Zhang}}]{Boyce2022}
---. 2022, \apj, 931, 7, \dodoi{10.3847/1538-4357/ac6104}

\bibitem[{{Bremer} {et~al.}(2011){Bremer}, {Witzel}, {Eckart}, {Zamaninasab}, {Buchholz}, {Sch{\"o}del}, {Straubmeier}, {Garc{\'\i}a-Mar{\'\i}n}, \& {Duschl}}]{Bremer2011}
{Bremer}, M., {Witzel}, G., {Eckart}, A., {et~al.} 2011, \aap, 532, A26, \dodoi{10.1051/0004-6361/201016134}

\bibitem[{{Brinkerink} {et~al.}(2021){Brinkerink}, {Falcke}, {Brunthaler}, \& {Law}}]{Brinkerink2021}
{Brinkerink}, C., {Falcke}, H., {Brunthaler}, A., \& {Law}, C. 2021, arXiv e-prints, arXiv:2107.13402, \dodoi{10.48550/arXiv.2107.13402}

\bibitem[{{Brinkerink} {et~al.}(2015){Brinkerink}, {Falcke}, {Law}, {Barkats}, {Bower}, {Brunthaler}, {Gammie}, {Impellizzeri}, {Markoff}, {Menten}, {Moscibrodzka}, {Peck}, {Rushton}, {Schaaf}, \& {Wright}}]{Brinkerink2015}
{Brinkerink}, C.~D., {Falcke}, H., {Law}, C.~J., {et~al.} 2015, \aap, 576, A41, \dodoi{10.1051/0004-6361/201424783}

\bibitem[{{Broderick} {et~al.}(2022){Broderick}, {Gold}, {Georgiev}, {Pesce}, {Tiede}, {Ni}, {Moriyama}, {Akiyama}, {Alberdi}, {Alef}, {Algaba}, {Anantua}, {Asada}, {Azulay}, {Bach}, {Baczko}, {Ball}, {Balokovi{\'c}}, {Barrett}, {Baub{\"o}ck}, {Benson}, {Bintley}, {Blackburn}, {Blundell}, {Bouman}, {Bower}, {Boyce}, {Bremer}, {Brinkerink}, {Brissenden}, {Britzen}, {Broguiere}, {Bronzwaer}, {Bustamante}, {Byun}, {Carlstrom}, {Ceccobello}, {Chael}, {Chan}, {Chatterjee}, {Chatterjee}, {Chen}, {Chen}, {Cheng}, {Cho}, {Christian}, {Conroy}, {Conway}, {Cordes}, {Crawford}, {Crew}, {Cruz-Osorio}, {Cui}, {Davelaar}, {De Laurentis}, {Deane}, {Dempsey}, {Desvignes}, {Dexter}, {Dhruv}, {Doeleman}, {Dougal}, {Dzib}, {Eatough}, {Emami}, {Falcke}, {Farah}, {Fish}, {Fomalont}, {Ford}, {Fraga-Encinas}, {Freeman}, {Friberg}, {Fromm}, {Fuentes}, {Galison}, {Gammie}, {Garc{\'\i}a}, {Gentaz}, {Goddi}, {G{\'o}mez-Ruiz}, {G{\'o}mez}, {Gu}, {Gurwell}, {Hada}, {Haggard}, {Haworth}, {Hecht}, {Hesper}, {Heumann}, {Ho}, {Ho}, {Honma},
  {Huang}, {Huang}, {Hughes}, {Ikeda}, {Impellizzeri}, {Inoue}, {Issaoun}, {James}, {Jannuzi}, {Janssen}, {Jeter}, {Jiang}, {Jim{\'e}nez-Rosales}, {Johnson}, {Jorstad}, {Joshi}, {Jung}, {Karami}, {Karuppusamy}, {Kawashima}, {Keating}, {Kettenis}, {Kim}, {Kim}, {Kim}, {Kim}, {Kino}, {Koay}, {Kocherlakota}, {Kofuji}, {Koch}, {Koyama}, {Kramer}, {Kramer}, {Krichbaum}, {Kuo}, {La Bella}, {Lauer}, {Lee}, {Lee}, {Leung}, {Levis}, {Li}, {Lico}, {Lindahl}, {Lindqvist}, {Lisakov}, {Liu}, {Liu}, {Liuzzo}, {Lo}, {Lobanov}, {Loinard}, {Lonsdale}, {Lu}, {Mao}, {Marchili}, {Markoff}, {Marrone}, {Marscher}, {Mart{\'\i}-Vidal}, {Matsushita}, {Matthews}, {Menten}, {Michalik}, {Mizuno}, {Mizuno}, {Moran}, {Moscibrodzka}, {M{\"u}ller}, {Mus}, {Musoke}, {Myserlis}, {Nadolski}, {Nagai}, {Nagar}, {Nakamura}, {Narayan}, {Narayanan}, {Natarajan}, {Nathanail}, {Navarro Fuentes}, {Neilsen}, {Neri}, {Noutsos}, {Nowak}, {Oh}, {Okino}, {Olivares}, {Ortiz-Le{\'o}n}, {Oyama}, {Palumbo}, {Paraschos}, {Park}, {Parsons}, {Patel}, {Pen},
  {Pi{\'e}tu}, {Plambeck}, {PopStefanija}, {Porth}, {P{\"o}tzl}, {Prather}, {Preciado-L{\'o}pez}, {Pu}, {Ramakrishnan}, {Rao}, {Rawlings}, {Raymond}, {Rezzolla}, {Ricarte}, {Ripperda}, {Roelofs}, {Rogers}, {Ros}, {Romero-Ca{\~n}izales}, {Roshanineshat}, {Rottmann}, {Roy}, {Ruiz}, {Ruszczyk}, {Rygl}, {S{\'a}nchez}, {S{\'a}nchez-Arg{\"u}elles}, {S{\'a}nchez-Portal}, {Sasada}, {Satapathy}, {Savolainen}, {Schloerb}, {Schonfeld}, {Schuster}, {Shao}, {Shen}, {Small}, {Sohn}, {SooHoo}, {Souccar}, {Sun}, {Tazaki}, {Tetarenko}, {Tilanus}, {Titus}, {Torne}, {Traianou}, {Trent}, {Trippe}, {Turk}, {van Bemmel}, {van Langevelde}, {van Rossum}, {Vos}, {Wagner}, {Ward-Thompson}, {Wardle}, {Weintroub}, {Wex}, {Wharton}, {Wielgus}, {Wiik}, {Witzel}, {Wondrak}, {Wong}, {Wu}, {Yamaguchi}, {Yoon}, {Young}, {Young}, {Younsi}, {Yuan}, {Yuan}, {Zensus}, {Zhao}, {Zhang}, \& {Zhao}}]{Broderick2022}
{Broderick}, A.~E., {Gold}, R., {Georgiev}, B., {et~al.} 2022, \apjl, 930, L21, \dodoi{10.3847/2041-8213/ac6584}

\bibitem[{{Capellupo} {et~al.}(2017){Capellupo}, {Haggard}, {Choux}, {Baganoff}, {Bower}, {Cotton}, {Degenaar}, {Dexter}, {Falcke}, {Fragile}, {Heinke}, {Law}, {Markoff}, {Neilsen}, {Ponti}, {Rea}, \& {Yusef-Zadeh}}]{Capellupo2017}
{Capellupo}, D.~M., {Haggard}, D., {Choux}, N., {et~al.} 2017, \apj, 845, 35, \dodoi{10.3847/1538-4357/aa7da6}

\bibitem[{{Chatterjee} \& {Narayan}(2022)}]{Chatterjee2022}
{Chatterjee}, K., \& {Narayan}, R. 2022, \apj, 941, 30, \dodoi{10.3847/1538-4357/ac9d97}

\bibitem[{{Chatterjee} {et~al.}(2021){Chatterjee}, {Markoff}, {Neilsen}, {Younsi}, {Witzel}, {Tchekhovskoy}, {Yoon}, {Ingram}, {van der Klis}, {Boyce}, {Do}, {Haggard}, \& {Nowak}}]{Chatterjee2021}
{Chatterjee}, K., {Markoff}, S., {Neilsen}, J., {et~al.} 2021, \mnras, 507, 5281, \dodoi{10.1093/mnras/stab2466}

\bibitem[{{Cheng} {et~al.}(2023){Cheng}, {Cho}, {Kawashima}, {Kino}, {Zhao}, {Algaba}, {Kofuji}, {Lee}, {Lee}, {Cheong}, {Jiang}, \& {Oh}}]{Cheng2023}
{Cheng}, X., {Cho}, I., {Kawashima}, T., {et~al.} 2023, Galaxies, 11, 46, \dodoi{10.3390/galaxies11020046}

\bibitem[{{Cho} {et~al.}(2022){Cho}, {Zhao}, {Kawashima}, {Kino}, {Akiyama}, {Johnson}, {Issaoun}, {Moriyama}, {Cheng}, {Algaba}, {Jung}, {Sohn}, {Krichbaum}, {Wielgus}, {Hada}, {Lu}, {Cui}, {Sawada-Satoh}, {Shen}, {Park}, {Jiang}, {Ro}, {Yi}, {Wajima}, {Lee}, {Hodgson}, {Tazaki}, {Honma}, {Niinuma}, {Trippe}, {An}, {Zhang}, {Lee}, {Oh}, {Byun}, {Lee}, {Kim}, {Oh}, {Koyama}, {Asada}, {Wang}, {Cui}, {Hagiwara}, {Nakamura}, {Takamura}, {Hirota}, {Sugiyama}, {Kawaguchi}, {Kobayashi}, {Oyama}, {Yonekura}, {Kim}, {Hwang}, {Jung}, {Kim}, {Kim}, {Oh}, {Roh}, {Yeom}, {Xia}, {Zhong}, {Li}, {Zhao}, {Wang}, {Liu}, \& {Chen}}]{Cho2022}
{Cho}, I., {Zhao}, G.-Y., {Kawashima}, T., {et~al.} 2022, \apj, 926, 108, \dodoi{10.3847/1538-4357/ac4165}

\bibitem[{{Condon} \& {Ransom}(2016)}]{CondonRansom2016}
{Condon}, J.~J., \& {Ransom}, S.~M. 2016, {Essential Radio Astronomy}

\bibitem[{{Dexter} {et~al.}(2014){Dexter}, {Kelly}, {Bower}, {Marrone}, {Stone}, \& {Plambeck}}]{Dexter2014}
{Dexter}, J., {Kelly}, B., {Bower}, G.~C., {et~al.} 2014, \mnras, 442, 2797, \dodoi{10.1093/mnras/stu1039}

\bibitem[{{Do} {et~al.}(2009){Do}, {Ghez}, {Morris}, {Yelda}, {Meyer}, {Lu}, {Hornstein}, \& {Matthews}}]{Do2009}
{Do}, T., {Ghez}, A.~M., {Morris}, M.~R., {et~al.} 2009, \apj, 691, 1021, \dodoi{10.1088/0004-637X/691/2/1021}

\bibitem[{{Do} {et~al.}(2019){Do}, {Witzel}, {Gautam}, {Chen}, {Ghez}, {Morris}, {Becklin}, {Ciurlo}, {Hosek}, {Martinez}, {Matthews}, {Sakai}, \& {Sch{\"o}del}}]{Do2019}
{Do}, T., {Witzel}, G., {Gautam}, A.~K., {et~al.} 2019, \apjl, 882, L27, \dodoi{10.3847/2041-8213/ab38c3}

\bibitem[{{Dodds-Eden} {et~al.}(2010){Dodds-Eden}, {Sharma}, {Quataert}, {Genzel}, {Gillessen}, {Eisenhauer}, \& {Porquet}}]{DE2010}
{Dodds-Eden}, K., {Sharma}, P., {Quataert}, E., {et~al.} 2010, \apj, 725, 450, \dodoi{10.1088/0004-637X/725/1/450}

\bibitem[{{Dodds-Eden} {et~al.}(2009){Dodds-Eden}, {Porquet}, {Trap}, {Quataert}, {Haubois}, {Gillessen}, {Grosso}, {Pantin}, {Falcke}, {Rouan}, {Genzel}, {Hasinger}, {Goldwurm}, {Yusef-Zadeh}, {Clenet}, {Trippe}, {Lagage}, {Bartko}, {Eisenhauer}, {Ott}, {Paumard}, {Perrin}, {Yuan}, {Fritz}, \& {Mascetti}}]{DoddsEden2009}
{Dodds-Eden}, K., {Porquet}, D., {Trap}, G., {et~al.} 2009, \apj, 698, 676, \dodoi{10.1088/0004-637X/698/1/676}

\bibitem[{{Eastman} {et~al.}(2010){Eastman}, {Siverd}, \& {Gaudi}}]{Eastman2010}
{Eastman}, J., {Siverd}, R., \& {Gaudi}, B.~S. 2010, \pasp, 122, 935, \dodoi{10.1086/655938}

\bibitem[{{Eckart} {et~al.}(2009){Eckart}, {Baganoff}, {Morris}, {Kunneriath}, {Zamaninasab}, {Witzel}, {Sch{\"o}del}, {Garc{\'\i}a-Mar{\'\i}n}, {Meyer}, {Bower}, {Marrone}, {Bautz}, {Brandt}, {Garmire}, {Ricker}, {Straubmeier}, {Roberts}, {Muzic}, {Mauerhan}, \& {Zensus}}]{Eckart2009}
{Eckart}, A., {Baganoff}, F.~K., {Morris}, M.~R., {et~al.} 2009, \aap, 500, 935, \dodoi{10.1051/0004-6361/200811354}

\bibitem[{{Evans} {et~al.}(2010){Evans}, {Primini}, {Glotfelty}, {Anderson}, {Bonaventura}, {Chen}, {Davis}, {Doe}, {Evans}, {Fabbiano}, {Galle}, {Gibbs}, {Grier}, {Hain}, {Hall}, {Harbo}, {He}, {Houck}, {Karovska}, {Kashyap}, {Lauer}, {McCollough}, {McDowell}, {Miller}, {Mitschang}, {Morgan}, {Mossman}, {Nichols}, {Nowak}, {Plummer}, {Refsdal}, {Rots}, {Siemiginowska}, {Sundheim}, {Tibbetts}, {Van Stone}, {Winkelman}, \& {Zografou}}]{Evans2010}
{Evans}, I.~N., {Primini}, F.~A., {Glotfelty}, K.~J., {et~al.} 2010, \apjs, 189, 37, \dodoi{10.1088/0067-0049/189/1/37}

\bibitem[{{Event Horizon Telescope Collaboration} {et~al.}(2022{\natexlab{a}}){Event Horizon Telescope Collaboration}, {Akiyama}, {Alberdi}, {Alef}, {Algaba}, {Anantua}, {Asada}, {Azulay}, {Bach}, {Baczko}, {Ball}, {Balokovi{\'c}}, {Barrett}, {Baub{\"o}ck}, {Benson}, {Bintley}, {Blackburn}, {Blundell}, {Bouman}, {Bower}, {Boyce}, {Bremer}, {Brinkerink}, {Brissenden}, {Britzen}, {Broderick}, {Broguiere}, {Bronzwaer}, {Bustamante}, {Byun}, {Carlstrom}, {Ceccobello}, {Chael}, {Chan}, {Chatterjee}, {Chatterjee}, {Chen}, {Chen}, {Cheng}, {Cho}, {Christian}, {Conroy}, {Conway}, {Cordes}, {Crawford}, {Crew}, {Cruz-Osorio}, {Cui}, {Davelaar}, {Laurentis}, {Deane}, {Dempsey}, {Desvignes}, {Dexter}, {Dhruv}, {Doeleman}, {Dougal}, {Dzib}, {Eatough}, {Emami}, {Falcke}, {Farah}, {Fish}, {Fomalont}, {Ford}, {Fraga-Encinas}, {Freeman}, {Friberg}, {Fromm}, {Fuentes}, {Galison}, {Gammie}, {Garc{\'\i}a}, {Gentaz}, {Georgiev}, {Goddi}, {Gold}, {G{\'o}mez-Ruiz}, {G{\'o}mez}, {Gu}, {Gurwell}, {Hada}, {Haggard}, {Haworth},
  {Hecht}, {Hesper}, {Heumann}, {Ho}, {Ho}, {Honma}, {Huang}, {Huang}, {Hughes}, {Ikeda}, {Impellizzeri}, {Inoue}, {Issaoun}, {James}, {Jannuzi}, {Janssen}, {Jeter}, {Jiang}, {Jim{\'e}nez-Rosales}, {Johnson}, {Jorstad}, {Joshi}, {Jung}, {Karami}, {Karuppusamy}, {Kawashima}, {Keating}, {Kettenis}, {Kim}, {Kim}, {Kim}, {Kim}, {Kino}, {Koay}, {Kocherlakota}, {Kofuji}, {Koch}, {Koyama}, {Kramer}, {Kramer}, {Krichbaum}, {Kuo}, {Bella}, {Lauer}, {Lee}, {Lee}, {Leung}, {Levis}, {Li}, {Lico}, {Lindahl}, {Lindqvist}, {Lisakov}, {Liu}, {Liu}, {Liuzzo}, {Lo}, {Lobanov}, {Loinard}, {Lonsdale}, {Lu}, {Mao}, {Marchili}, {Markoff}, {Marrone}, {Marscher}, {Mart{\'\i}-Vidal}, {Matsushita}, {Matthews}, {Medeiros}, {Menten}, {Michalik}, {Mizuno}, {Mizuno}, {Moran}, {Moriyama}, {Moscibrodzka}, {M{\"u}ller}, {Mus}, {Musoke}, {Myserlis}, {Nadolski}, {Nagai}, {Nagar}, {Nakamura}, {Narayan}, {Narayanan}, {Natarajan}, {Nathanail}, {Fuentes}, {Neilsen}, {Neri}, {Ni}, {Noutsos}, {Nowak}, {Oh}, {Okino}, {Olivares}, {Ortiz-Le{\'o}n},
  {Oyama}, {{\"O}zel}, {Palumbo}, {Paraschos}, {Park}, {Parsons}, {Patel}, {Pen}, {Pesce}, {Pi{\'e}tu}, {Plambeck}, {PopStefanija}, {Porth}, {P{\"o}tzl}, {Prather}, {Preciado-L{\'o}pez}, {Psaltis}, {Pu}, {Ramakrishnan}, {Rao}, {Rawlings}, {Raymond}, {Rezzolla}, {Ricarte}, {Ripperda}, {Roelofs}, {Rogers}, {Ros}, {Romero-Ca{\~n}izales}, {Roshanineshat}, {Rottmann}, {Roy}, {Ruiz}, {Ruszczyk}, {Rygl}, {S{\'a}nchez}, {S{\'a}nchez-Arg{\"u}elles}, {S{\'a}nchez-Portal}, {Sasada}, {Satapathy}, {Savolainen}, {Schloerb}, {Schonfeld}, {Schuster}, {Shao}, {Shen}, {Small}, {Sohn}, {SooHoo}, {Souccar}, {Sun}, {Tazaki}, {Tetarenko}, {Tiede}, {Tilanus}, {Titus}, {Torne}, {Traianou}, {Trent}, {Trippe}, {Turk}, {van Bemmel}, {van Langevelde}, {van Rossum}, {Vos}, {Wagner}, {Ward-Thompson}, {Wardle}, {Weintroub}, {Wex}, {Wharton}, {Wielgus}, {Wiik}, {Witzel}, {Wondrak}, {Wong}, {Wu}, {Yamaguchi}, {Yoon}, {Young}, {Young}, {Younsi}, {Yuan}, {Yuan}, {Zensus}, {Zhang}, {Zhao}, {Zhao}, {Agurto}, {Allardi}, {Amestica}, {Araneda},
  {Arriagada}, {Berghuis}, {Bertarini}, {Berthold}, {Blanchard}, {Brown}, {C{\'a}rdenas}, {Cantzler}, {Caro}, {Castillo-Dom{\'\i}nguez}, {Chan}, {Chang}, {Chang}, {Chang}, {Chang}, {Chen}, {Chilson}, {Chuter}, {Ciechanowicz}, {Colin-Beltran}, {Coulson}, {Crowley}, {Degenaar}, {Dornbusch}, {Dur{\'a}n}, {Everett}, {Faber}, {Forster}, {Fuchs}, {Gale}, {Geertsema}, {Gonz{\'a}lez}, {Graham}, {Gueth}, {Halverson}, {Han}, {Han}, {Hasegawa}, {Hern{\'a}ndez-Rebollar}, {Herrera}, {Herrero-Illana}, {Heyminck}, {Hirota}, {Hoge}, {Hostler Schimpf}, {Howie}, {Huang}, {Jiang}, {Jinchi}, {John}, {Kimura}, {Klein}, {Kubo}, {Kuroda}, {Kwon}, {Lacasse}, {Laing}, {Leitch}, {Li}, {Liu}, {Liu}, {Lin}, {Lu}, {Mac-Auliffe}, {Martin-Cocher}, {Matulonis}, {Maute}, {Messias}, {Meyer-Zhao}, {Monta{\~n}a}, {Montenegro-Montes}, {Montgomerie}, {Moreno Nolasco}, {Muders}, {Nishioka}, {Norton}, {Nystrom}, {Ogawa}, {Olivares}, {Oshiro}, {P{\'e}rez-Beaupuits}, {Parra}, {Phillips}, {Poirier}, {Pradel}, {Qiu}, {Raffin}, {Rahlin}, {Ram{\'\i}rez},
  {Ressler}, {Reynolds}, {Rodr{\'\i}guez-Montoya}, {Saez-Madain}, {Santana}, {Shaw}, {Shirkey}, {Silva}, {Snow}, {Sousa}, {Sridharan}, {Stahm}, {Stark}, {Test}, {Torstensson}, {Venegas}, {Walther}, {Wei}, {White}, {Wieching}, {Wijnands}, {Wouterloot}, {Yu}, {Yu (于威)}, \& {Zeballos}}]{EHT2022a}
{Event Horizon Telescope Collaboration}, {Akiyama}, K., {Alberdi}, A., {et~al.} 2022{\natexlab{a}}, \apjl, 930, L12, \dodoi{10.3847/2041-8213/ac6674}

\bibitem[{{Event Horizon Telescope Collaboration} {et~al.}(2022{\natexlab{b}}){Event Horizon Telescope Collaboration}, {Akiyama}, {Alberdi}, {Alef}, {Algaba}, {Anantua}, {Asada}, {Azulay}, {Bach}, {Baczko}, {Ball}, {Balokovi{\'c}}, {Barrett}, {Baub{\"o}ck}, {Benson}, {Bintley}, {Blackburn}, {Blundell}, {Bouman}, {Bower}, {Boyce}, {Bremer}, {Brinkerink}, {Brissenden}, {Britzen}, {Broderick}, {Broguiere}, {Bronzwaer}, {Bustamante}, {Byun}, {Carlstrom}, {Ceccobello}, {Chael}, {Chan}, {Chatterjee}, {Chatterjee}, {Chen}, {Chen}, {Cheng}, {Cho}, {Christian}, {Conroy}, {Conway}, {Cordes}, {Crawford}, {Crew}, {Cruz-Osorio}, {Cui}, {Davelaar}, {De Laurentis}, {Deane}, {Dempsey}, {Desvignes}, {Dexter}, {Dhruv}, {Doeleman}, {Dougal}, {Dzib}, {Eatough}, {Emami}, {Falcke}, {Farah}, {Fish}, {Fomalont}, {Ford}, {Fraga-Encinas}, {Freeman}, {Friberg}, {Fromm}, {Fuentes}, {Galison}, {Gammie}, {Garc{\'\i}a}, {Gentaz}, {Georgiev}, {Goddi}, {Gold}, {G{\'o}mez-Ruiz}, {G{\'o}mez}, {Gu}, {Gurwell}, {Hada}, {Haggard}, {Haworth},
  {Hecht}, {Hesper}, {Heumann}, {Ho}, {Ho}, {Honma}, {Huang}, {Huang}, {Hughes}, {Ikeda}, {Violette Impellizzeri}, {Inoue}, {Issaoun}, {James}, {Jannuzi}, {Janssen}, {Jeter}, {Jiang}, {Jim{\'e}nez-Rosales}, {Johnson}, {Jorstad}, {Joshi}, {Jung}, {Karami}, {Karuppusamy}, {Kawashima}, {Keating}, {Kettenis}, {Kim}, {Kim}, {Kim}, {Kim}, {Kino}, {Koay}, {Kocherlakota}, {Kofuji}, {Koch}, {Koyama}, {Kramer}, {Kramer}, {Krichbaum}, {Kuo}, {Bella}, {Lauer}, {Lee}, {Lee}, {Leung}, {Levis}, {Li}, {Lico}, {Lindahl}, {Lindqvist}, {Lisakov}, {Liu}, {Liu}, {Liuzzo}, {Lo}, {Lobanov}, {Loinard}, {Lonsdale}, {Lu}, {Mao}, {Marchili}, {Markoff}, {Marrone}, {Marscher}, {Mart{\'\i}-Vidal}, {Matsushita}, {Matthews}, {Medeiros}, {Menten}, {Michalik}, {Mizuno}, {Mizuno}, {Moran}, {Moriyama}, {Moscibrodzka}, {M{\"u}ller}, {Mus}, {Musoke}, {Myserlis}, {Nadolski}, {Nagai}, {Nagar}, {Nakamura}, {Narayan}, {Narayanan}, {Natarajan}, {Nathanail}, {Navarro Fuentes}, {Neilsen}, {Neri}, {Ni}, {Noutsos}, {Nowak}, {Oh}, {Okino}, {Olivares},
  {Ortiz-Le{\'o}n}, {Oyama}, {{\"O}zel}, {Palumbo}, {Filippos Paraschos}, {Park}, {Parsons}, {Patel}, {Pen}, {Pesce}, {Pi{\'e}tu}, {Plambeck}, {PopStefanija}, {Porth}, {P{\"o}tzl}, {Prather}, {Preciado-L{\'o}pez}, {Psaltis}, {Pu}, {Ramakrishnan}, {Rao}, {Rawlings}, {Raymond}, {Rezzolla}, {Ricarte}, {Ripperda}, {Roelofs}, {Rogers}, {Ros}, {Romero-Ca{\~n}izales}, {Roshanineshat}, {Rottmann}, {Roy}, {Ruiz}, {Ruszczyk}, {Rygl}, {S{\'a}nchez}, {S{\'a}nchez-Arg{\"u}elles}, {S{\'a}nchez-Portal}, {Sasada}, {Satapathy}, {Savolainen}, {Schloerb}, {Schonfeld}, {Schuster}, {Shao}, {Shen}, {Small}, {Sohn}, {SooHoo}, {Souccar}, {Sun}, {Tazaki}, {Tetarenko}, {Tiede}, {Tilanus}, {Titus}, {Torne}, {Traianou}, {Trent}, {Trippe}, {Turk}, {van Bemmel}, {van Langevelde}, {van Rossum}, {Vos}, {Wagner}, {Ward-Thompson}, {Wardle}, {Weintroub}, {Wex}, {Wharton}, {Wielgus}, {Wiik}, {Witzel}, {Wondrak}, {Wong}, {Wu}, {Yamaguchi}, {Yoon}, {Young}, {Young}, {Younsi}, {Yuan}, {Yuan}, {Zensus}, {Zhang}, {Zhao}, {Zhao}, {Chan}, {Qiu},
  {Ressler}, \& {White}}]{EHT2022e}
---. 2022{\natexlab{b}}, \apjl, 930, L16, \dodoi{10.3847/2041-8213/ac6672}

\bibitem[{{Fazio} {et~al.}(2018){Fazio}, {Hora}, {Witzel}, {Willner}, {Ashby}, {Baganoff}, {Becklin}, {Carey}, {Haggard}, {Gammie}, {Ghez}, {Gurwell}, {Ingalls}, {Marrone}, {Morris}, \& {Smith}}]{Fazio2018}
{Fazio}, G.~G., {Hora}, J.~L., {Witzel}, G., {et~al.} 2018, \apj, 864, 58, \dodoi{10.3847/1538-4357/aad4a2}

\bibitem[{{Fragione} \& {Loeb}(2020)}]{Fragione2020}
{Fragione}, G., \& {Loeb}, A. 2020, \apjl, 901, L32, \dodoi{10.3847/2041-8213/abb9b4}

\bibitem[{{Fragione} \& {Loeb}(2022)}]{Fragione2022}
---. 2022, \apjl, 932, L17, \dodoi{10.3847/2041-8213/ac76ca}

\bibitem[{{Fritz} {et~al.}(2011){Fritz}, {Gillessen}, {Dodds-Eden}, {Lutz}, {Genzel}, {Raab}, {Ott}, {Pfuhl}, {Eisenhauer}, \& {Yusef-Zadeh}}]{Fritz2011}
{Fritz}, T.~K., {Gillessen}, S., {Dodds-Eden}, K., {et~al.} 2011, \apj, 737, 73, \dodoi{10.1088/0004-637X/737/2/73}

\bibitem[{{Fruscione} {et~al.}(2006){Fruscione}, {McDowell}, {Allen}, {Brickhouse}, {Burke}, {Davis}, {Durham}, {Elvis}, {Galle}, {Harris}, {Huenemoerder}, {Houck}, {Ishibashi}, {Karovska}, {Nicastro}, {Noble}, {Nowak}, {Primini}, {Siemiginowska}, {Smith}, \& {Wise}}]{Ciao2006}
{Fruscione}, A., {McDowell}, J.~C., {Allen}, G.~E., {et~al.} 2006, in Society of Photo-Optical Instrumentation Engineers (SPIE) Conference Series, Vol. 6270, Society of Photo-Optical Instrumentation Engineers (SPIE) Conference Series, ed. D.~R. {Silva} \& R.~E. {Doxsey}, 62701V, \dodoi{10.1117/12.671760}

\bibitem[{{Gravity Collaboration} {et~al.}(2018){Gravity Collaboration}, {Abuter}, {Amorim}, {Baub{\"o}ck}, {Berger}, {Bonnet}, {Brandner}, {Cl{\'e}net}, {Coud{\'e} Du Foresto}, {de Zeeuw}, {Deen}, {Dexter}, {Duvert}, {Eckart}, {Eisenhauer}, {F{\"o}rster Schreiber}, {Garcia}, {Gao}, {Gendron}, {Genzel}, {Gillessen}, {Guajardo}, {Habibi}, {Haubois}, {Henning}, {Hippler}, {Horrobin}, {Huber}, {Jim{\'e}nez-Rosales}, {Jocou}, {Kervella}, {Lacour}, {Lapeyr{\`e}re}, {Lazareff}, {Le Bouquin}, {L{\'e}na}, {Lippa}, {Ott}, {Panduro}, {Paumard}, {Perraut}, {Perrin}, {Pfuhl}, {Plewa}, {Rabien}, {Rodr{\'\i}guez-Coira}, {Rousset}, {Sternberg}, {Straub}, {Straubmeier}, {Sturm}, {Tacconi}, {Vincent}, {von Fellenberg}, {Waisberg}, {Widmann}, {Wieprecht}, {Wiezorrek}, {Woillez}, \& {Yazici}}]{Gravity2018}
{Gravity Collaboration}, {Abuter}, R., {Amorim}, A., {et~al.} 2018, \aap, 618, L10, \dodoi{10.1051/0004-6361/201834294}

\bibitem[{{Gravity Collaboration} {et~al.}(2019){Gravity Collaboration}, {Abuter}, {Amorim}, {Baub{\"o}ck}, {Berger}, {Bonnet}, {Brandner}, {Cl{\'e}net}, {Coud{\'e} Du Foresto}, {de Zeeuw}, {Dexter}, {Duvert}, {Eckart}, {Eisenhauer}, {F{\"o}rster Schreiber}, {Garcia}, {Gao}, {Gendron}, {Genzel}, {Gerhard}, {Gillessen}, {Habibi}, {Haubois}, {Henning}, {Hippler}, {Horrobin}, {Jim{\'e}nez-Rosales}, {Jocou}, {Kervella}, {Lacour}, {Lapeyr{\`e}re}, {Le Bouquin}, {L{\'e}na}, {Ott}, {Paumard}, {Perraut}, {Perrin}, {Pfuhl}, {Rabien}, {Rodriguez Coira}, {Rousset}, {Scheithauer}, {Sternberg}, {Straub}, {Straubmeier}, {Sturm}, {Tacconi}, {Vincent}, {von Fellenberg}, {Waisberg}, {Widmann}, {Wieprecht}, {Wiezorrek}, {Woillez}, \& {Yazici}}]{Gravity2019}
---. 2019, \aap, 625, L10, \dodoi{10.1051/0004-6361/201935656}

\bibitem[{{GRAVITY Collaboration} {et~al.}(2020{\natexlab{a}}){GRAVITY Collaboration}, {Jim{\'e}nez-Rosales}, {Dexter}, {Widmann}, {Baub{\"o}ck}, {Abuter}, {Amorim}, {Berger}, {Bonnet}, {Brandner}, {Cl{\'e}net}, {de Zeeuw}, {Eckart}, {Eisenhauer}, {F{\"o}rster Schreiber}, {Garcia}, {Gao}, {Gendron}, {Genzel}, {Gillessen}, {Habibi}, {Haubois}, {Hei{\ss}el}, {Henning}, {Hippler}, {Horrobin}, {Jochum}, {Jocou}, {Kaufer}, {Kervella}, {Lacour}, {Lapeyr{\`e}re}, {Le Bouquin}, {L{\'e}na}, {Nowak}, {Ott}, {Paumard}, {Perraut}, {Perrin}, {Pfuhl}, {Rodr{\'\i}guez-Coira}, {Shangguan}, {Scheithauer}, {Stadler}, {Straub}, {Straubmeier}, {Sturm}, {Tacconi}, {Vincent}, {von Fellenberg}, {Waisberg}, {Wieprecht}, {Wiezorrek}, {Woillez}, {Yazici}, \& {Zins}}]{Gravity2020}
{GRAVITY Collaboration}, {Jim{\'e}nez-Rosales}, A., {Dexter}, J., {et~al.} 2020{\natexlab{a}}, \aap, 643, A56, \dodoi{10.1051/0004-6361/202038283}

\bibitem[{{GRAVITY Collaboration} {et~al.}(2020{\natexlab{b}}){GRAVITY Collaboration}, {Baub{\"o}ck}, {Dexter}, {Abuter}, {Amorim}, {Berger}, {Bonnet}, {Brandner}, {Cl{\'e}net}, {Coud{\'e} Du Foresto}, {de Zeeuw}, {Duvert}, {Eckart}, {Eisenhauer}, {F{\"o}rster Schreiber}, {Gao}, {Garcia}, {Gendron}, {Genzel}, {Gerhard}, {Gillessen}, {Habibi}, {Haubois}, {Henning}, {Hippler}, {Horrobin}, {Jim{\'e}nez-Rosales}, {Jocou}, {Kervella}, {Lacour}, {Lapeyr{\`e}re}, {Le Bouquin}, {L{\'e}na}, {Ott}, {Paumard}, {Perraut}, {Perrin}, {Pfuhl}, {Rabien}, {Rodriguez Coira}, {Rousset}, {Scheithauer}, {Stadler}, {Sternberg}, {Straub}, {Straubmeier}, {Sturm}, {Tacconi}, {Vincent}, {von Fellenberg}, {Waisberg}, {Widmann}, {Wieprecht}, {Wiezorrek}, {Woillez}, \& {Yazici}}]{Gravity2020orbit}
{GRAVITY Collaboration}, {Baub{\"o}ck}, M., {Dexter}, J., {et~al.} 2020{\natexlab{b}}, \aap, 635, A143, \dodoi{10.1051/0004-6361/201937233}

\bibitem[{{GRAVITY Collaboration} {et~al.}(2021){GRAVITY Collaboration}, {Abuter}, {Amorim}, {Baub{\"o}ck}, {Baganoff}, {Berger}, {Boyce}, {Bonnet}, {Brandner}, {Cl{\'e}net}, {Davies}, {de Zeeuw}, {Dexter}, {Dallilar}, {Drescher}, {Eckart}, {Eisenhauer}, {Fazio}, {F{\"o}rster Schreiber}, {Foster}, {Gammie}, {Garcia}, {Gao}, {Gendron}, {Genzel}, {Ghisellini}, {Gillessen}, {Gurwell}, {Habibi}, {Haggard}, {Hailey}, {Harrison}, {Haubois}, {Hei{\ss}el}, {Henning}, {Hippler}, {Hora}, {Horrobin}, {Jim{\'e}nez-Rosales}, {Jochum}, {Jocou}, {Kaufer}, {Kervella}, {Lacour}, {Lapeyr{\`e}re}, {Le Bouquin}, {L{\'e}na}, {Lowrance}, {Lutz}, {Markoff}, {Mori}, {Morris}, {Neilsen}, {Nowak}, {Ott}, {Paumard}, {Perraut}, {Perrin}, {Ponti}, {Pfuhl}, {Rabien}, {Rodr{\'\i}guez-Coira}, {Shangguan}, {Shimizu}, {Scheithauer}, {Smith}, {Stadler}, {Stern}, {Straub}, {Straubmeier}, {Sturm}, {Tacconi}, {Vincent}, {von Fellenberg}, {Waisberg}, {Widmann}, {Wieprecht}, {Wiezorrek}, {Willner}, {Witzel}, {Woillez}, {Yazici}, {Young}, {Zhang},
  \& {Zins}}]{Abuter2021}
{GRAVITY Collaboration}, {Abuter}, R., {Amorim}, A., {et~al.} 2021, \aap, 654, A22, \dodoi{10.1051/0004-6361/202140981}

\bibitem[{{Grillmair} {et~al.}(2014){Grillmair}, {Carey}, {Stauffer}, \& {Ingalls}}]{Grillmair2014}
{Grillmair}, C.~J., {Carey}, S.~J., {Stauffer}, J.~R., \& {Ingalls}, J.~G. 2014, in Society of Photo-Optical Instrumentation Engineers (SPIE) Conference Series, Vol. 9143, Space Telescopes and Instrumentation 2014: Optical, Infrared, and Millimeter Wave, ed. J.~{Oschmann}, Jacobus~M., M.~{Clampin}, G.~G. {Fazio}, \& H.~A. {MacEwen}, 914359, \dodoi{10.1117/12.2057238}

\bibitem[{{Grillmair} {et~al.}(2012){Grillmair}, {Carey}, {Stauffer}, {Fisher}, {Olds}, {Ingalls}, {Krick}, {Glaccum}, {Laine}, {Lowrance}, \& {Surace}}]{Grillmair2012}
{Grillmair}, C.~J., {Carey}, S.~J., {Stauffer}, J.~R., {et~al.} 2012, in Society of Photo-Optical Instrumentation Engineers (SPIE) Conference Series, Vol. 8448, Observatory Operations: Strategies, Processes, and Systems IV, ed. A.~B. {Peck}, R.~L. {Seaman}, \& F.~{Comeron}, 84481I, \dodoi{10.1117/12.927191}

\bibitem[{Harris {et~al.}(2020)Harris, Millman, van~der Walt, Gommers, Virtanen, Cournapeau, Wieser, Taylor, Berg, Smith, Kern, Picus, Hoyer, van Kerkwijk, Brett, Haldane, del R{\'{i}}o, Wiebe, Peterson, G{\'{e}}rard-Marchant, Sheppard, Reddy, Weckesser, Abbasi, Gohlke, \& Oliphant}]{numpy}
Harris, C.~R., Millman, K.~J., van~der Walt, S.~J., {et~al.} 2020, Nature, 585, 357, \dodoi{10.1038/s41586-020-2649-2}

\bibitem[{{Herrnstein} {et~al.}(2004){Herrnstein}, {Zhao}, {Bower}, \& {Goss}}]{Herrnstein2004}
{Herrnstein}, R.~M., {Zhao}, J.-H., {Bower}, G.~C., \& {Goss}, W.~M. 2004, \aj, 127, 3399, \dodoi{10.1086/420711}

\bibitem[{{Hora} {et~al.}(2014){Hora}, {Witzel}, {Ashby}, {Becklin}, {Carey}, {Fazio}, {Ghez}, {Ingalls}, {Meyer}, {Morris}, {Smith}, \& {Willner}}]{Hora2014}
{Hora}, J.~L., {Witzel}, G., {Ashby}, M.~L.~N., {et~al.} 2014, \apj, 793, 120, \dodoi{10.1088/0004-637X/793/2/120}

\bibitem[{Hunter(2007)}]{matplotlib}
Hunter, J.~D. 2007, Computing in Science \& Engineering, 9, 90, \dodoi{10.1109/MCSE.2007.55}

\bibitem[{{Iwata} {et~al.}(2020){Iwata}, {Oka}, {Tsuboi}, {Miyoshi}, \& {Takekawa}}]{Iwata2020}
{Iwata}, Y., {Oka}, T., {Tsuboi}, M., {Miyoshi}, M., \& {Takekawa}, S. 2020, \apjl, 892, L30, \dodoi{10.3847/2041-8213/ab800d}

\bibitem[{{Lu} {et~al.}(2011){Lu}, {Krichbaum}, {Eckart}, {K{\"o}nig}, {Kunneriath}, {Witzel}, {Witzel}, \& {Zensus}}]{Lu2011}
{Lu}, R.~S., {Krichbaum}, T.~P., {Eckart}, A., {et~al.} 2011, \aap, 525, A76, \dodoi{10.1051/0004-6361/200913807}

\bibitem[{{Markoff} {et~al.}(2001){Markoff}, {Falcke}, {Yuan}, \& {Biermann}}]{Markoff2001}
{Markoff}, S., {Falcke}, H., {Yuan}, F., \& {Biermann}, P.~L. 2001, \aap, 379, L13, \dodoi{10.1051/0004-6361:20011346}

\bibitem[{{Marrone} {et~al.}(2008){Marrone}, {Baganoff}, {Morris}, {Moran}, {Ghez}, {Hornstein}, {Dowell}, {Mu{\~n}oz}, {Bautz}, {Ricker}, {Brandt}, {Garmire}, {Lu}, {Matthews}, {Zhao}, {Rao}, \& {Bower}}]{Marrone2008}
{Marrone}, D.~P., {Baganoff}, F.~K., {Morris}, M.~R., {et~al.} 2008, \apj, 682, 373, \dodoi{10.1086/588806}

\bibitem[{{M}c{K}inney(2010)}]{pandas}
{M}c{K}inney, W. 2010, in {P}roceedings of the 9th {P}ython in {S}cience {C}onference, ed. {S}t\'efan van~der {W}alt \& {J}arrod {M}illman, 56 -- 61, \dodoi{10.25080/Majora-92bf1922-00a}

\bibitem[{{McMullin} {et~al.}(2007){McMullin}, {Waters}, {Schiebel}, {Young}, \& {Golap}}]{McMullin2007}
{McMullin}, J.~P., {Waters}, B., {Schiebel}, D., {Young}, W., \& {Golap}, K. 2007, in Astronomical Society of the Pacific Conference Series, Vol. 376, Astronomical Data Analysis Software and Systems XVI, ed. R.~A. {Shaw}, F.~{Hill}, \& D.~J. {Bell}, 127

\bibitem[{{Meyer} {et~al.}(2008){Meyer}, {Do}, {Ghez}, {Morris}, {Witzel}, {Eckart}, {B{\'e}langer}, \& {Sch{\"o}del}}]{Meyer2008}
{Meyer}, L., {Do}, T., {Ghez}, A., {et~al.} 2008, \apjl, 688, L17, \dodoi{10.1086/593147}

\bibitem[{{Michail} {et~al.}(2021{\natexlab{a}}){Michail}, {Wardle}, {Yusef-Zadeh}, \& {Kunneriath}}]{Michail2021c}
{Michail}, J.~M., {Wardle}, M., {Yusef-Zadeh}, F., \& {Kunneriath}, D. 2021{\natexlab{a}}, \apj, 923, 54, \dodoi{10.3847/1538-4357/ac2d2c}

\bibitem[{{Michail} {et~al.}(2021{\natexlab{b}}){Michail}, {Yusef-Zadeh}, \& {Wardle}}]{Michail2021b}
{Michail}, J.~M., {Yusef-Zadeh}, F., \& {Wardle}, M. 2021{\natexlab{b}}, \mnras, 505, 3616, \dodoi{10.1093/mnras/stab1529}

\bibitem[{{Michail} {et~al.}(2023){Michail}, {Yusef-Zadeh}, {Wardle}, \& {Kunneriath}}]{Michail2023}
{Michail}, J.~M., {Yusef-Zadeh}, F., {Wardle}, M., \& {Kunneriath}, D. 2023, \mnras, 520, 2644, \dodoi{10.1093/mnras/stad291}

\bibitem[{{Mossoux} \& {Grosso}(2017)}]{Mossoux2017}
{Mossoux}, E., \& {Grosso}, N. 2017, \aap, 604, A85, \dodoi{10.1051/0004-6361/201629778}

\bibitem[{{Mossoux} {et~al.}(2016){Mossoux}, {Grosso}, {Bushouse}, {Eckart}, {Yusef-Zadeh}, {Plambeck}, {Peissker}, {Valencia-S.}, {Porquet}, {Cotton}, \& {Roberts}}]{Mossoux2016}
{Mossoux}, E., {Grosso}, N., {Bushouse}, H., {et~al.} 2016, \aap, 589, A116, \dodoi{10.1051/0004-6361/201527554}

\bibitem[{{Murchikova} \& {Witzel}(2021)}]{Murchikova2021}
{Murchikova}, L., \& {Witzel}, G. 2021, \apjl, 920, L7, \dodoi{10.3847/2041-8213/ac2308}

\bibitem[{{Nathanail} {et~al.}(2020){Nathanail}, {Fromm}, {Porth}, {Olivares}, {Younsi}, {Mizuno}, \& {Rezzolla}}]{Nathanail2020}
{Nathanail}, A., {Fromm}, C.~M., {Porth}, O., {et~al.} 2020, \mnras, 495, 1549, \dodoi{10.1093/mnras/staa1165}

\bibitem[{{Neilsen} {et~al.}(2013){Neilsen}, {Nowak}, {Gammie}, {Dexter}, {Markoff}, {Haggard}, {Nayakshin}, {Wang}, {Grosso}, {Porquet}, {Tomsick}, {Degenaar}, {Fragile}, {Houck}, {Wijnands}, {Miller}, \& {Baganoff}}]{Neilsen2013}
{Neilsen}, J., {Nowak}, M.~A., {Gammie}, C., {et~al.} 2013, \apj, 774, 42, \dodoi{10.1088/0004-637X/774/1/42}

\bibitem[{{Neilsen} {et~al.}(2015){Neilsen}, {Markoff}, {Nowak}, {Dexter}, {Witzel}, {Barri{\`e}re}, {Li}, {Baganoff}, {Degenaar}, {Fragile}, {Gammie}, {Goldwurm}, {Grosso}, \& {Haggard}}]{Neilsen2015}
{Neilsen}, J., {Markoff}, S., {Nowak}, M.~A., {et~al.} 2015, \apj, 799, 199, \dodoi{10.1088/0004-637X/799/2/199}

\bibitem[{Newville {et~al.}(2023)Newville, Otten, Nelson, Stensitzki, Ingargiola, Allan, Fox, Carter, Michał, Osborn, Pustakhod, lneuhaus, Weigand, Aristov, Glenn, Deil, mgunyho, Mark, Hansen, Pasquevich, Foks, Zobrist, Frost, Stuermer, azelcer, Polloreno, Persaud, Nielsen, Pompili, \& Eendebak}]{LMFIT}
Newville, M., Otten, R., Nelson, A., {et~al.} 2023, lmfit/lmfit-py: 1.2.2, 1.2.2,  Zenodo, \dodoi{10.5281/zenodo.8145703}

\bibitem[{{Nowak} {et~al.}(2012){Nowak}, {Neilsen}, {Markoff}, {Baganoff}, {Porquet}, {Grosso}, {Levin}, {Houck}, {Eckart}, {Falcke}, {Ji}, {Miller}, \& {Wang}}]{Nowak2012}
{Nowak}, M.~A., {Neilsen}, J., {Markoff}, S.~B., {et~al.} 2012, \apj, 759, 95, \dodoi{10.1088/0004-637X/759/2/95}

\bibitem[{Pedregosa {et~al.}(2011)Pedregosa, Varoquaux, Gramfort, Michel, Thirion, Grisel, Blondel, Prettenhofer, Weiss, Dubourg, Vanderplas, Passos, Cournapeau, Brucher, Perrot, \& Duchesnay}]{sklearn}
Pedregosa, F., Varoquaux, G., Gramfort, A., {et~al.} 2011, Journal of Machine Learning Research, 12, 2825

\bibitem[{{Peterson} {et~al.}(1998){Peterson}, {Wanders}, {Horne}, {Collier}, {Alexander}, {Kaspi}, \& {Maoz}}]{Peterson1998}
{Peterson}, B.~M., {Wanders}, I., {Horne}, K., {et~al.} 1998, \pasp, 110, 660, \dodoi{10.1086/316177}

\bibitem[{{Ponti} {et~al.}(2015){Ponti}, {De Marco}, {Morris}, {Merloni}, {Mu{\~n}oz-Darias}, {Clavel}, {Haggard}, {Zhang}, {Nandra}, {Gillessen}, {Mori}, {Neilsen}, {Rea}, {Degenaar}, {Terrier}, \& {Goldwurm}}]{Ponti2015}
{Ponti}, G., {De Marco}, B., {Morris}, M.~R., {et~al.} 2015, \mnras, 454, 1525, \dodoi{10.1093/mnras/stv1537}

\bibitem[{{Ponti} {et~al.}(2017){Ponti}, {George}, {Scaringi}, {Zhang}, {Jin}, {Dexter}, {Terrier}, {Clavel}, {Degenaar}, {Eisenhauer}, {Genzel}, {Gillessen}, {Goldwurm}, {Habibi}, {Haggard}, {Hailey}, {Harrison}, {Merloni}, {Mori}, {Nandra}, {Ott}, {Pfuhl}, {Plewa}, \& {Waisberg}}]{Ponti2017}
{Ponti}, G., {George}, E., {Scaringi}, S., {et~al.} 2017, \mnras, 468, 2447, \dodoi{10.1093/mnras/stx596}

\bibitem[{{Porth} {et~al.}(2021){Porth}, {Mizuno}, {Younsi}, \& {Fromm}}]{Porth2021}
{Porth}, O., {Mizuno}, Y., {Younsi}, Z., \& {Fromm}, C.~M. 2021, \mnras, 502, 2023, \dodoi{10.1093/mnras/stab163}

\bibitem[{{Pringle}(1981)}]{Pringle1981}
{Pringle}, J.~E. 1981, \araa, 19, 137, \dodoi{10.1146/annurev.aa.19.090181.001033}

\bibitem[{{Ressler} {et~al.}(2020{\natexlab{a}}){Ressler}, {Quataert}, \& {Stone}}]{Ressler2020a}
{Ressler}, S.~M., {Quataert}, E., \& {Stone}, J.~M. 2020{\natexlab{a}}, \mnras, 492, 3272, \dodoi{10.1093/mnras/stz3605}

\bibitem[{{Ressler} {et~al.}(2020{\natexlab{b}}){Ressler}, {White}, {Quataert}, \& {Stone}}]{Ressler2020b}
{Ressler}, S.~M., {White}, C.~J., {Quataert}, E., \& {Stone}, J.~M. 2020{\natexlab{b}}, \apjl, 896, L6, \dodoi{10.3847/2041-8213/ab9532}

\bibitem[{{Ripperda} {et~al.}(2020){Ripperda}, {Bacchini}, \& {Philippov}}]{Ripperda2020}
{Ripperda}, B., {Bacchini}, F., \& {Philippov}, A.~A. 2020, \apj, 900, 100, \dodoi{10.3847/1538-4357/ababab}

\bibitem[{{Scargle}(1998)}]{Scargle1998}
{Scargle}, J.~D. 1998, \apj, 504, 405, \dodoi{10.1086/306064}

\bibitem[{{Scargle} {et~al.}(2013){Scargle}, {Norris}, {Jackson}, \& {Chiang}}]{Scargle2013}
{Scargle}, J.~D., {Norris}, J.~P., {Jackson}, B., \& {Chiang}, J. 2013, \apj, 764, 167, \dodoi{10.1088/0004-637X/764/2/167}

\bibitem[{{Subroweit} {et~al.}(2017){Subroweit}, {Garc{\'\i}a-Mar{\'\i}n}, {Eckart}, {Borkar}, {Valencia-S.}, {Witzel}, {Shahzamanian}, \& {Straubmeier}}]{Subroweit2017}
{Subroweit}, M., {Garc{\'\i}a-Mar{\'\i}n}, M., {Eckart}, A., {et~al.} 2017, \aap, 601, A80, \dodoi{10.1051/0004-6361/201628530}

\bibitem[{{Sun} {et~al.}(2018){Sun}, {Grier}, \& {Peterson}}]{PYCCF}
{Sun}, M., {Grier}, C.~J., \& {Peterson}, B.~M. 2018, {PyCCF: Python Cross Correlation Function for reverberation mapping studies}.
\newblock \doeprint{1805.032}

\bibitem[{{van der Laan}(1966)}]{VDL1966}
{van der Laan}, H. 1966, \nat, 211, 1131, \dodoi{10.1038/2111131a0}

\bibitem[{Virtanen {et~al.}(2020)Virtanen, Gommers, Oliphant, Haberland, Reddy, Cournapeau, Burovski, Peterson, Weckesser, Bright, {van der Walt}, Brett, Wilson, Millman, Mayorov, Nelson, Jones, Kern, Larson, Carey, Polat, Feng, Moore, {VanderPlas}, Laxalde, Perktold, Cimrman, Henriksen, Quintero, Harris, Archibald, Ribeiro, Pedregosa, {van Mulbregt}, \& {SciPy 1.0 Contributors}}]{scipy}
Virtanen, P., Gommers, R., Oliphant, T.~E., {et~al.} 2020, Nature Methods, 17, 261, \dodoi{10.1038/s41592-019-0686-2}

\bibitem[{{Weldon} {et~al.}(2023){Weldon}, {Do}, {Witzel}, {Ghez}, {Gautam}, {Becklin}, {Morris}, {Martinez}, {Sakai}, {Lu}, {Matthews}, {Hosek}, \& {Haggard}}]{Weldon2023}
{Weldon}, G.~C., {Do}, T., {Witzel}, G., {et~al.} 2023, arXiv e-prints, arXiv:2308.09749, \dodoi{10.48550/arXiv.2308.09749}

\bibitem[{{Wielgus} {et~al.}(2022){Wielgus}, {Marchili}, {Mart{\'\i}-Vidal}, {Keating}, {Ramakrishnan}, {Tiede}, {Fomalont}, {Issaoun}, {Neilsen}, {Nowak}, {Blackburn}, {Gammie}, {Goddi}, {Haggard}, {Lee}, {Moscibrodzka}, {Tetarenko}, {Bower}, {Chan}, {Chatterjee}, {Chesler}, {Dexter}, {Doeleman}, {Georgiev}, {Gurwell}, {Johnson}, {Marrone}, {Mus}, {Psaltis}, {Ripperda}, {Witzel}, {Akiyama}, {Alberdi}, {Alef}, {Algaba}, {Anantua}, {Asada}, {Azulay}, {Bach}, {Baczko}, {Ball}, {Balokovi{\'c}}, {Barrett}, {Baub{\"o}ck}, {Benson}, {Bintley}, {Blundell}, {Boland}, {Bouman}, {Boyce}, {Bremer}, {Brinkerink}, {Brissenden}, {Britzen}, {Broderick}, {Broguiere}, {Bronzwaer}, {Bustamante}, {Byun}, {Carlstrom}, {Ceccobello}, {Chael}, {Chatterjee}, {Chen}, {Chen}, {Cho}, {Christian}, {Conroy}, {Conway}, {Cordes}, {Crawford}, {Crew}, {Cruz-Osorio}, {Cui}, {Davelaar}, {De Laurentis}, {Deane}, {Dempsey}, {Desvignes}, {Dhruv}, {Dzib}, {Eatough}, {Emami}, {Falcke}, {Farah}, {Fish}, {Ford}, {Fraga-Encinas}, {Freeman}, {Friberg},
  {Fromm}, {Fuentes}, {Galison}, {Garc{\'\i}a}, {Gentaz}, {Gold}, {G{\'o}mez-Ruiz}, {G{\'o}mez}, {Gu}, {Hada}, {Haworth}, {Hecht}, {Hesper}, {Ho}, {Ho}, {Honma}, {Huang}, {Huang}, {Hughes}, {Ikeda}, {Impellizzeri}, {Inoue}, {James}, {Jannuzi}, {Janssen}, {Jeter}, {Jiang}, {Jim{\'e}nez-Rosales}, {Jorstad}, {Joshi}, {Jung}, {Karami}, {Karuppusamy}, {Kawashima}, {Kettenis}, {Kim}, {Kim}, {Kim}, {Kim}, {Kino}, {Koay}, {Kocherlakota}, {Kofuji}, {Koch}, {Koyama}, {Kramer}, {Kramer}, {Krichbaum}, {Kuo}, {La Bella}, {Lauer}, {Lee}, {Leung}, {Levis}, {Li}, {Lico}, {Lindahl}, {Lindqvist}, {Lisakov}, {Liu}, {Liu}, {Liuzzo}, {Lo}, {Lobanov}, {Loinard}, {Lonsdale}, {Lu}, {Mao}, {Markoff}, {Marscher}, {Matsushita}, {Matthews}, {Medeiros}, {Menten}, {Michalik}, {Mizuno}, {Mizuno}, {Moran}, {Moriyama}, {M{\"u}ller}, {Musoke}, {Myserlis}, {Nadolski}, {Nagai}, {Nagar}, {Nakamura}, {Narayan}, {Narayanan}, {Natarajan}, {Nathanail}, {Navarro Fuentes}, {Neri}, {Ni}, {Noutsos}, {Oh}, {Okino}, {Olivares}, {Ortiz-Le{\'o}n}, {Oyama},
  {{\"O}zel}, {Palumbo}, {Paraschos}, {Park}, {Parsons}, {Patel}, {Pen}, {Pesce}, {Pi{\'e}tu}, {Plambeck}, {PopStefanija}, {Porth}, {P{\"o}tzl}, {Prather}, {Preciado-L{\'o}pez}, {Pu}, {Rao}, {Rawlings}, {Raymond}, {Rezzolla}, {Ricarte}, {Roelofs}, {Rogers}, {Ros}, {Romero-Canizales}, {Roshanineshat}, {Rottmann}, {Roy}, {Ruiz}, {Ruszczyk}, {Rygl}, {S{\'a}nchez}, {S{\'a}nchez-Arg{\"u}elles}, {S{\'a}nchez-Portal}, {Sasada}, {Satapathy}, {Savolainen}, {Schloerb}, {Schuster}, {Shao}, {Shen}, {Small}, {Won Sohn}, {SooHoo}, {Souccar}, {Sun}, {Tazaki}, {Tilanus}, {Titus}, {Torne}, {Traianou}, {Trent}, {Trippe}, {van Bemmel}, {van Langevelde}, {van Rossum}, {Vos}, {Wagner}, {Ward-Thompson}, {Wardle}, {Weintroub}, {Wex}, {Wharton}, {Wiik}, {Wondrak}, {Wong}, {Wu}, {Yamaguchi}, {Yoon}, {Young}, {Young}, {Younsi}, {Yuan}, {Yuan}, {Zensus}, {Zhang}, {Zhao}, \& {Zhao}}]{Wielgus2022a}
{Wielgus}, M., {Marchili}, N., {Mart{\'\i}-Vidal}, I., {et~al.} 2022, \apjl, 930, L19, \dodoi{10.3847/2041-8213/ac6428}

\bibitem[{{Williams} {et~al.}(2017){Williams}, {Clavel}, {Newton}, \& {Ryzhkov}}]{BBlocks}
{Williams}, P. K.~G., {Clavel}, M., {Newton}, E., \& {Ryzhkov}, D. 2017, {pwkit: Astronomical utilities in Python}.
\newblock \doeprint{1704.001}

\bibitem[{{Witzel} {et~al.}(2012){Witzel}, {Eckart}, {Bremer}, {Zamaninasab}, {Shahzamanian}, {Valencia-S.}, {Sch{\"o}del}, {Karas}, {Lenzen}, {Marchili}, {Sabha}, {Garcia-Marin}, {Buchholz}, {Kunneriath}, \& {Straubmeier}}]{Witzel2012}
{Witzel}, G., {Eckart}, A., {Bremer}, M., {et~al.} 2012, \apjs, 203, 18, \dodoi{10.1088/0067-0049/203/2/18}

\bibitem[{{Witzel} {et~al.}(2018){Witzel}, {Martinez}, {Hora}, {Willner}, {Morris}, {Gammie}, {Becklin}, {Ashby}, {Baganoff}, {Carey}, {Do}, {Fazio}, {Ghez}, {Glaccum}, {Haggard}, {Herrero-Illana}, {Ingalls}, {Narayan}, \& {Smith}}]{Witzel2018}
{Witzel}, G., {Martinez}, G., {Hora}, J., {et~al.} 2018, \apj, 863, 15, \dodoi{10.3847/1538-4357/aace62}

\bibitem[{{Witzel} {et~al.}(2021){Witzel}, {Martinez}, {Willner}, {Becklin}, {Boyce}, {Do}, {Eckart}, {Fazio}, {Ghez}, {Gurwell}, {Haggard}, {Herrero-Illana}, {Hora}, {Li}, {Liu}, {Marchili}, {Morris}, {Smith}, {Subroweit}, \& {Zensus}}]{Witzel2021}
{Witzel}, G., {Martinez}, G., {Willner}, S.~P., {et~al.} 2021, \apj, 917, 73, \dodoi{10.3847/1538-4357/ac0891}

\bibitem[{{Yuan} \& {Wang}(2016)}]{Yuan2016}
{Yuan}, Q., \& {Wang}, Q.~D. 2016, \mnras, 456, 1438, \dodoi{10.1093/mnras/stv2778}

\bibitem[{{Yusef-Zadeh} {et~al.}(2006{\natexlab{a}}){Yusef-Zadeh}, {Roberts}, {Wardle}, {Heinke}, \& {Bower}}]{FYZ2006}
{Yusef-Zadeh}, F., {Roberts}, D., {Wardle}, M., {Heinke}, C.~O., \& {Bower}, G.~C. 2006{\natexlab{a}}, \apj, 650, 189, \dodoi{10.1086/506375}

\bibitem[{{Yusef-Zadeh} {et~al.}(2010){Yusef-Zadeh}, {Wardle}, {Bushouse}, {Dowell}, \& {Roberts}}]{FYZ2010}
{Yusef-Zadeh}, F., {Wardle}, M., {Bushouse}, H., {Dowell}, C.~D., \& {Roberts}, D.~A. 2010, \apjl, 724, L9, \dodoi{10.1088/2041-8205/724/1/L9}

\bibitem[{{Yusef-Zadeh} {et~al.}(2008){Yusef-Zadeh}, {Wardle}, {Heinke}, {Dowell}, {Roberts}, {Baganoff}, \& {Cotton}}]{FYZ2008}
{Yusef-Zadeh}, F., {Wardle}, M., {Heinke}, C., {et~al.} 2008, \apj, 682, 361, \dodoi{10.1086/588803}

\bibitem[{{Yusef-Zadeh} {et~al.}(2011){Yusef-Zadeh}, {Wardle}, {Miller-Jones}, {Roberts}, {Grosso}, \& {Porquet}}]{FYZ2011}
{Yusef-Zadeh}, F., {Wardle}, M., {Miller-Jones}, J.~C.~A., {et~al.} 2011, \apj, 729, 44, \dodoi{10.1088/0004-637X/729/1/44}

\bibitem[{{Yusef-Zadeh} {et~al.}(2006{\natexlab{b}}){Yusef-Zadeh}, {Bushouse}, {Dowell}, {Wardle}, {Roberts}, {Heinke}, {Bower}, {Vila-Vilar{\'o}}, {Shapiro}, {Goldwurm}, \& {B{\'e}langer}}]{YusefZadeh2006a}
{Yusef-Zadeh}, F., {Bushouse}, H., {Dowell}, C.~D., {et~al.} 2006{\natexlab{b}}, \apj, 644, 198, \dodoi{10.1086/503287}

\bibitem[{{Yusef-Zadeh} {et~al.}(2006{\natexlab{c}}){Yusef-Zadeh}, {Bushouse}, {Dowell}, {Wardle}, {Roberts}, {Heinke}, {Bower}, {Vila-Vilar{\'o}}, {Shapiro}, {Goldwurm}, \& {B{\'e}langer}}]{FYZ2006a}
---. 2006{\natexlab{c}}, \apj, 644, 198, \dodoi{10.1086/503287}

\bibitem[{{Yusef-Zadeh} {et~al.}(2009){Yusef-Zadeh}, {Bushouse}, {Wardle}, {Heinke}, {Roberts}, {Dowell}, {Brunthaler}, {Reid}, {Martin}, {Marrone}, {Porquet}, {Grosso}, {Dodds-Eden}, {Bower}, {Wiesemeyer}, {Miyazaki}, {Pal}, {Gillessen}, {Goldwurm}, {Trap}, \& {Maness}}]{FYZ2009}
{Yusef-Zadeh}, F., {Bushouse}, H., {Wardle}, M., {et~al.} 2009, \apj, 706, 348, \dodoi{10.1088/0004-637X/706/1/348}

\end{thebibliography}
\bibliographystyle{aasjournal}



\end{document}